\documentclass[aps,pra,reprint,longbibliography]{revtex4-1}
\usepackage{graphicx}% Include figure files
\usepackage{dcolumn}% Align table columns on decimal point
\usepackage{bm}% bold math
\usepackage{xcolor}
\begin{document}

\title{Selective Excitation of Subwavelength Atomic Clouds}
\author{Rasoul Alaee{$^{*1,2}$}, Akbar Safari$^{1,3}$, and Robert W. Boyd{$^{1,4}$}}
% Authors and affiliations
\address{
$^{1}$Department of Physics, University of Ottawa, 25 Templeton, Ottawa, Ontario K1N 6N5, Canada\\
$^{2}$Karlsruhe Institute of Technology, Institute of Theoretical Solid State Physics, Wolfgang-Gaede-Str. 1,
D-76131 Karlsruhe, Germany\\
$^{3}$Department of Physics, University of Wisconsin-Madison, Madison, Wisconsin 53706, USA\\
$^{4}$The Institute of Optics, University of Rochester, Rochester, New York 14627, USA\\
$*$ $\rm{Corresponding\,\, author: rasoul.alaee@gmail.com}$
}
%%\date{\today}

\begin{abstract}
A dense cloud of atoms with randomly changing positions exhibits coherent and incoherent scattering. We show that an atomic cloud of subwavelength dimensions can be modeled as a single scatterer where both coherent and incoherent components of the scattered photons can be fully explained based on \textit{effective} multipole moments. This model allows us to arrive at a relation between the coherent and incoherent components of scattering based on the conservation of energy. Furthermore, using superposition of four plane waves, we show that one can \textit{selectively} excite different multipole moments and thus tailor the scattering of the atomic cloud to control the cooperative shift, resonance linewidth, and the radiation pattern. Our approach provides a new insight into the scattering phenomena in atomic ensembles and opens a pathway towards controlling scattering for applications such as generation and manipulation of single-photon states.
\end{abstract}
                          
\maketitle
\section{Introduction}
Since Dicke's original work in 1954, the physics of collective effects and multiple scattering of light by a dense ensemble of atoms has attracted significant attention~\cite{Dicke:1954,Lehmberg1970,Friedberg1973,Haroche1982,Guerin:2017}. In particular, remarkable phenomena such as Anderson localization~\cite{Kramer1993,Skipetrov2015magnetic}, coherent backscattering~\cite{Labeyrie1999}, random lasing~\cite{Baudouin2013}, superradiance~\cite{Dicke:1954,Scully2009,Bienaim:2012,Araujo2016superradiance}, subradiance~\cite{Dicke:1954,Bienaim:2012,Guerin2016subradiance}, and cooperative shift~\cite{Roof2016} have been explored for cold ensembles of atoms. The physical origin of these phenomena can be understood by multiple scattering of light in a collection of atoms~\cite{Sheng2006Book}. An ideal platform for observation of these cooperative effects is an array of cold atoms with subwavelength distances~\cite{Jenkins:2012,Jenkins:2013,meir2014Lambshift,Bettles2015,Bettles2016Cooperative,Bettles2016,Facchinetti2016storing,Asenjo2017,shahmoon2017,Barredo:2018,wild2018quantum,Facchinetti2018,Zoller2019,Bekenstein2020,Rui:2020,Alaee2020,Alaee:2020Kerker,Ballantine2020,Solntsev2020Reveiw,Andreoli:2021}. However, arranging atoms in \textit{arbitrary subwavelength} structures is highly demanding and cannot be achieved easily~\cite{Rui:2020}. On the other hand, it has been demonstrated that a cloud of cold atoms can reach densities with atomic distances less than the resonant wavelength where a strong coherent dipole-dipole interaction
couples the atoms~\cite{Pellegrino:2014,Jennewein:2016}. Therefore, the atoms interact with light collectively~\cite{Pellegrino:2014,Jennewein:2016,Corman:2017,Schilder:2016,Schilder:2017,Browaeys2020}. Nonetheless, the linewidth and frequency of each collective mode depends strongly on the exact spatial arrangement of the atoms, which changes randomly even in a cold ensemble of atoms. As a consequence, the atomic cloud exhibits both coherent and incoherent scattering~\cite{Schilder:2016,Schilder:2017}. Moreover, the random motion of the atoms seems to weaken the cooperative effects significantly and causes a subwavelength cloud of atoms to scatter fewer photons on average compared to a single atom, in contrast to Dicke's work~\cite{Schilder:2020}. 

In this paper, we show that the cooperative shift and resonance linewidth of a subwavelength cloud of cold atoms can be controlled by structuring the excitation field. Structured light beams enable properties and applications in both classical and quantum optics~\cite{Andrews2011Book,Dunlop2016, ChekhovaBanzer2021book}. In particular, structured light offers unique control of many phenomena including angstrom localization and detection of nanoparticles~\cite{Roy2015, Neugebauer2016,Zheng2016,Bag2018}, Kerker effects and directional scattering~\cite{Xi2016broadband, Wei2017, wozniak2015selective},  counter-intuitive optical pulling and lateral forces~\cite{chen2011optical, sukhov2017}, and nonlinear microscopy~\cite{Bautista2016}, among other feats~\cite{Andrews2011Book,das2015beam,Dunlop2016,ChekhovaBanzer2021book}. However, the potential of structured light to manipulate cooperative effects remains unexplored.

We introduce a multipolar decomposition and demonstrate that both coherent and \textit{incoherent} scattering of a subwavelength atomic cloud can be fully characterized by electric and magnetic multipole moments. Using conservation of energy and multipolar decomposition, we find analytical expressions that relate fluctuating and averaged electric and magnetic polarizabilities. Then, by employing superposition of four plane waves, we selectively excite the electric and magnetic multipole moments. As a result, we can control the cooperative shift and resonance linewidth of the atomic cloud by changing the relative phase between the plane waves.

%%%%%%%%%%%%%% Figure 1 %%%%%%%%%%%%%%
\begin{figure*}
    \centering
    \includegraphics[width=0.99\textwidth]{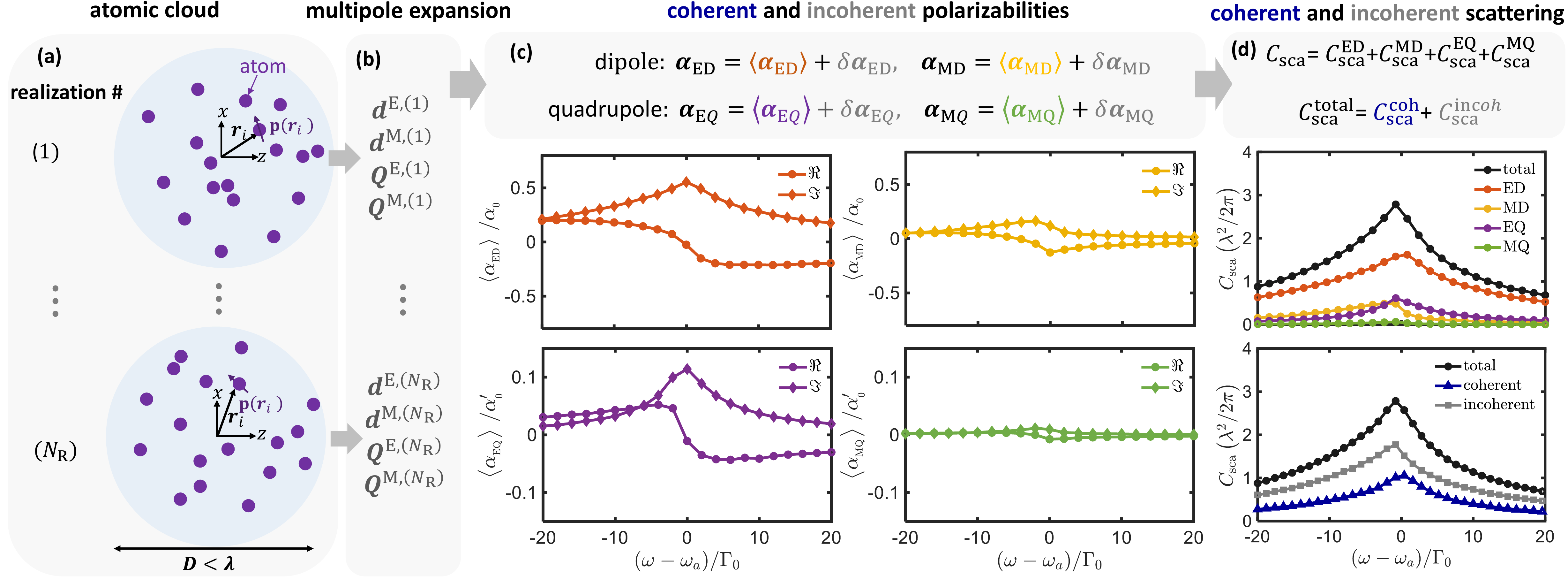}
    \caption{(a) Schematic drawing of a subwavelength atomic cloud composed of $N$ atoms at random positions for different realizations. $N_{\rm R}$ is the number of realizations. The diameter $D$ of the atomic cloud is smaller than the wavelength. (b) The induced electric and magnetic multipole moments of each realization obtained from multipole expansion of the induced current [using Eqs.~(\ref{eq:ME_Exact}) in the Appendix]. (c) Coherent electric and magnetic polarizabilities as a function of frequency detuning obtained from the induced multipole moments shown in (b). (d) Coherent and incoherent scattering cross sections and contribution of each multipole moment obtained from Eq.~(\ref{eq:C_sca_total_1PW_MT}). The ensemble-averages are obtained from 10000 realizations of the atomic cloud with radius $R=0.2\lambda_a$ composed of $N=25$ atoms.}
    \label{fig:Average_polarizability}
\end{figure*}

%%%%%%%%%%%%%%%%%%%%%%%%%%%%%%%%%%%%%%
\section{Weak excitation limit}
We consider a subwavelength cloud of atoms uniformly distributed in a sphere with radius R~[see Fig.~\ref{fig:Average_polarizability}(a)]. The atomic cloud is assumed to be dense, i.e.,  $\rho/k^{3}>1$, where $\rho=N/V$ is the spatial density, $N$ is the number of atoms, and $V$ is the volume of the atomic cloud. We assume cold atoms without nonradiative losses and with a negligible Doppler effect compared to their radiative linewidth as in experimental realizations~\cite{Pellegrino:2014,Jennewein:2016}. The atomic cloud is investigated in the weak excitation limit such that the atomic transition is far below the saturation limit. Thus, each atom in the cloud is modeled by an isotropic electric polarizability given by $\alpha\left(\omega\right)=-(\alpha_{0}\Gamma_{0}/2)/\left(\omega-\omega_{a}+i\Gamma_{0}/2\right)$, where $\Gamma_{0}$ is the radiative linewidth, $\omega_{a}$ is the atomic transition angular frequency, and  $\omega-\omega_{a}\ll\omega_{a}$ is the detuning of the illumination from the atomic resonance. $\alpha_{0}=6\pi/k^{3}$, where $k=\omega/c$ is the wavenumber of the illumination~\cite{lagendijk1996,Lagendijk1998,lambropoulos2007,Alaee:2017Review}.

\section{Coherent and incoherent multipole expansion}
We assume that the atoms in the subwavelength cloud have random spatial distributions. We consider many realizations for which the position of the atoms are changed with a uniform probability distribution~[see Fig.~\ref{fig:Average_polarizability}(a)]. The atomic cloud is illuminated by plane waves and the total scattered field can be decomposed into two parts: $\ensuremath{\mathbf{E}_{{\rm sca}}=\left\langle \mathbf{E}_{{\rm sca}}\right\rangle +\mathbf{\delta E}_{{\rm sca}}}$, where $\left\langle \mathbf{E}_{{\rm sca}}\right\rangle$ and $\mathbf{\delta E}_{{\rm sca}}$ are the coherent (ensemble-averaged) and incoherent (fluctuating) fields, respectively~\cite{Schilder:2016,Schilder:2017,Schilder:2020}. The induced polarization current density of the atomic cloud is given by $\mathbf{J}\left(\mathbf{r},\omega\right)  =  -i\omega\sum_{i=1}^{N}\mathbf{p}\left(\mathbf{r}_{i}\right)\delta\left(\mathbf{r}-\mathbf{r}_{i}\right)$~\cite{tai1994dyadic,Jackson1999,Alaee2020,Alaee:2020Kerker}, where $\delta$ is the Dirac delta function and $\mathbf{p}\left(\mathbf{r}_{i}\right)$
is the induced electric dipole moment of the $i$th atom placed at $\mathbf{r}_{i}$~[see Fig.~\ref{fig:Average_polarizability}(a)]. Now by employing multipole decomposition of the current $\mathbf{J}\left(\mathbf{r},\omega\right)$~\cite{Alaee:2018,Alaee2019}, we can calculate the induced \textit{effective} multipole moments of the atomic cloud which can be decomposed into coherent and incoherent parts~(see Appendix~\ref{APP_ME} for details):
\begin{eqnarray}
d^{\rm E}_{\mu} & = & \left\langle d^{\rm E}_{\mu}\right\rangle +\delta d^{\rm E}_{\mu},\,\,\,\,\,\,\,\,\,\,d^{\rm M}_{\mu}=\left\langle d^{\rm M}_{\mu}\right\rangle +\delta d^{\rm M}_{\mu},\nonumber \\
Q^{\rm E}_{\mu\nu} & = & \left\langle Q^{\rm E}_{\mu\nu}\right\rangle +\delta Q^{\rm E}_{\mu\nu},\,\,\,\,\,\,\,\,\,\,Q^{\rm M}_{\mu\nu}=\left\langle Q^{\rm M}_{\mu\nu}\right\rangle +\delta Q^{\rm M}_{\mu\nu},\label{eq:ME_coh_incoh}
\end{eqnarray}
where, $\mu,\nu \in {x,y,z}$. The quantities $d^{\rm E}_{\mu}$, $d^{\rm M}_{\mu}$, $Q^{\rm E}_{\mu\nu}$, and $Q^{\rm M}_{\mu\nu}$ are the effective
electric dipole (ED), magnetic dipole (MD), electric quadrupole (EQ), and magnetic quadrupole (MQ) moments of the atomic cloud, respectively. The symbol $\left\langle \, \right\rangle$ represents an ensemble-average. Using the induced multipole moments, we obtain the electric and magnetic dipole and quadrupole polarizabilities\begin{equation}
\bm{\alpha}_{i}=\left\langle \bm{\alpha}_{i}\right\rangle +\delta\bm{\alpha}_{i},\,\,\,\,\,i\in\left\{ \mathrm{ED,MD,EQ,MQ}\right\}, 
\end{equation}where $\left\langle \bm{\alpha}_{i}\right\rangle$ and $\delta\bm{\alpha}_{i}$ are the coherent and incoherent polarizabilities, respectively. 

%%%%%%%%%%%%%% Figure 2 %%%%%%%%%%%%%%
\begin{figure*}
    \centering
    \includegraphics[width=0.7\textwidth]{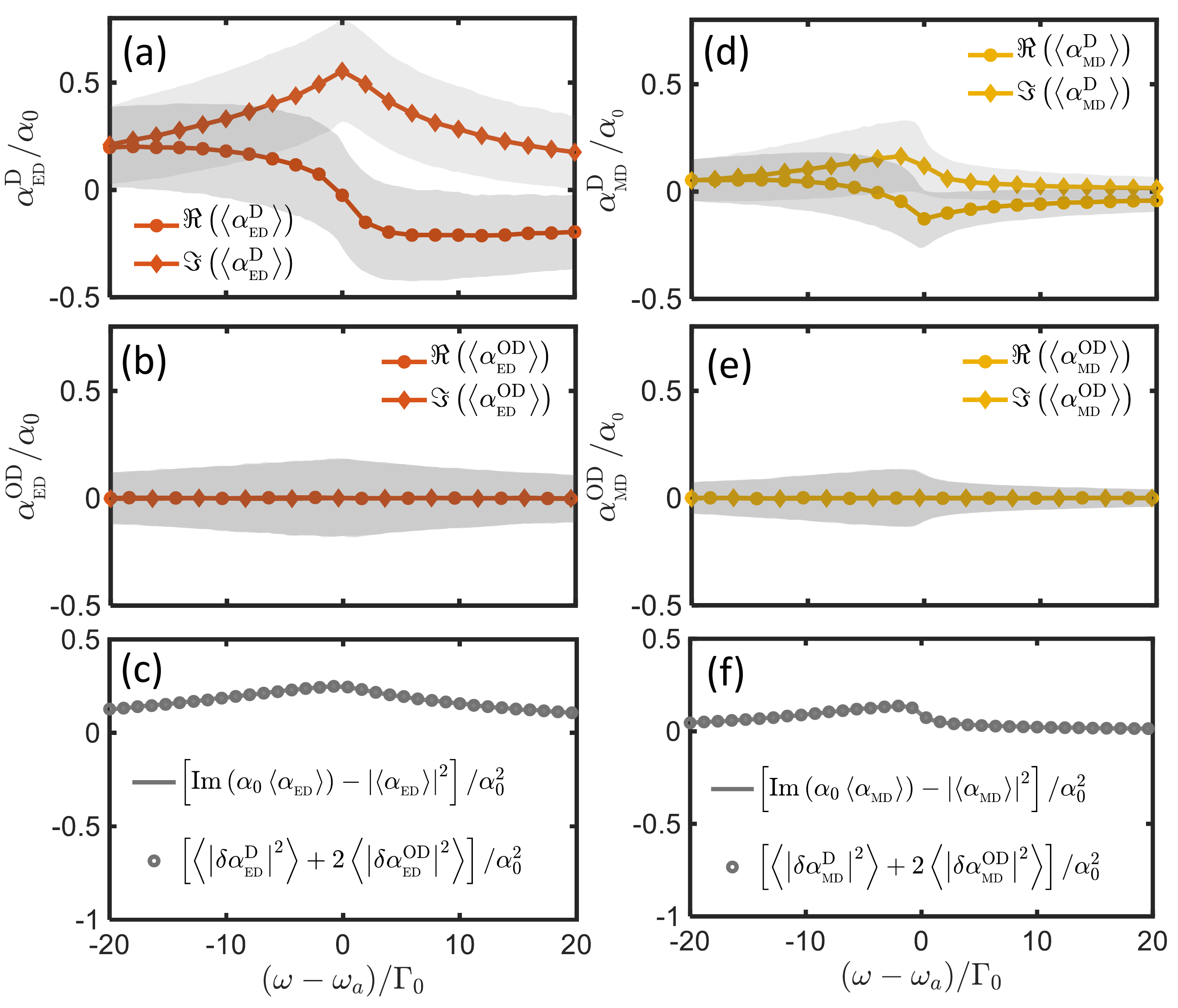}
    \caption{Coherent and incoherent electric and magnetic polarizabilities obtained from the induced multipole moments and defined as $\alpha_{i}^{j}=\left\langle \alpha_{i}^{j}\right\rangle +\delta\alpha_{i}^{j}$, where $i\in\left\{ \mathrm{ED,MD}\right\} ,j\in\left\{ \mathrm{D,OD}\right\}$. ED (MD) denotes the electric (magnetic) dipole and D (OD) represents diagonal (off-diagonal) components of the polarizability tensor (see Eq.~\ref{polarizability_D_OD}). (a)-(b) Electric dipole polarizabilities as a function of frequency detuning. Note that $\left\langle \alpha_{\mathrm{_{ED}}}^{\mathrm{D}}\right\rangle \neq0$ whereas $\left\langle \alpha_{_{\mathrm{ED}}}^{\mathrm{OD}}\right\rangle =0$. The thickness of the shaded lines show the fluctuating components of the polarizabilities. (c) Relation between coherent and incoherent polarizabilities obtained from the conservation of energy, see the left and right sides of Eq.~(\ref{eq:OT_ED}). (d)-(f) Same as (a)-(c) for the magnetic dipole polarizability of the atomic cloud.}
    \label{fig:polarizability_Coh_incoh}
\end{figure*}

%%%%%%%%%%%%%%%%%%%%%%%%%%%%%%%%%%%%%%  

\section{Single plane wave excitation} We consider a subwavelength atomic cloud composed of $N$ atoms as shown in Fig.~\ref{fig:Average_polarizability} (a) and illuminated by an $x$-polarized plane wave $\mathbf{E_{\rm inc}}=E_{0}e^{ikz}\mathbf{e}_{x}$ propagating in the $z$ direction. The ensemble-averaged induced multipole moments are given by~(see Appendix~\ref{APP_1PW} for details)
\begin{eqnarray}
\left\langle \mathbf{d}^{\mathrm{E}}\right\rangle  & = & \varepsilon_{0}\left\langle \alpha_{_{\mathrm{ED}}}\right\rangle E_{0}\mathbf{e}_{x},\,\,\,\,\,\,\left\langle \mathbf{d}^{\mathrm{M}}\right\rangle =\left\langle \alpha_{_{\mathrm{MD}}}\right\rangle H_{0} \mathbf{e}_{y},\nonumber\\
\left\langle \mathbf{Q}^{\mathrm{E}}\right\rangle  & = & \frac{ik}{2}\varepsilon_{0}\left\langle \alpha_{_{\mathrm{EQ}}}\right\rangle E_{0}\left(\mathbf{e}_{x}\mathbf{e}_{z}+\mathbf{e}_{z}\mathbf{e}_{x}\right),\nonumber\\
\left\langle \mathbf{Q}^{\mathrm{M}}\right\rangle  & = & \frac{ik}{2}\left\langle \alpha_{_{\mathrm{MQ}}}\right\rangle H_{0}\left(\mathbf{e}_{y}\mathbf{e}_{z}+\mathbf{e}_{z}\mathbf{e}_{y}\right),\label{eq:dQ_1PW}
\end{eqnarray}
 where, $\mathbf{Q}^{\mathrm{E}}$ and $\mathbf{Q}^{\mathrm{M}}$ are tensors of rank two, $\mathbf{e}_{\mu}\mathbf{e}_{\nu}$ is the unit dyad, $\mu,\nu \in {x,y,z}$, and $H_0$ is the amplitude of the  magnetic field of the plane wave. Having the polarizabilities, we can calculate the coherent and total scattering cross sections by~(see Appendix~\ref{APP_1PW})
 \begin{eqnarray}
 C_{\mathrm{sca}}^{\mathrm{coh}}&=&\frac{3\lambda^{2}}{2\pi\alpha_0^2}\left(\left|\left\langle \alpha_{_{\mathrm{ED}}}\right\rangle \right|^{2}+\left|\left\langle \alpha_{_{\mathrm{MD}}}\right\rangle \right|^{2}\right)\nonumber\\
 & & + \frac{5\lambda^{2}}{2\pi{\alpha_0^{\prime}}^{2}}\left(\left|\left\langle \alpha_{_{\mathrm{EQ}}}\right\rangle \right|^{2}+\left|\left\langle \alpha_{_{\mathrm{MQ}}}\right\rangle \right|^{2}\right),\nonumber\\
 C_{\mathrm{sca}}^{\mathrm{\mathrm{total}}}&=&\frac{3\lambda^{2}}{2\pi\alpha_0^2}\mathrm{Im}\left(\left\langle \alpha_{_{\mathrm{ED}}}\right\rangle +\left\langle \alpha_{_{\mathrm{MD}}}\right\rangle \right)\nonumber\\
 & &+\frac{5\lambda^{2}}{2\pi{\alpha_0^{\prime}}^2}\mathrm{Im}\left(\left\langle \alpha_{_{\mathrm{EQ}}}\right\rangle +\left\langle \alpha_{_{\mathrm{MQ}}}\right\rangle \right),\label{eq:C_sca_total_1PW_MT}
 \end{eqnarray}
where $\alpha_0 = 6\pi/k^3$ ($\alpha_0^{\prime} = 120\pi/k^5$) is related to the radiation loss of a dipole (quadrupole) moment. Equation (\ref{eq:C_sca_total_1PW_MT}), which is the first main result of this paper, allows us to calculate the coherent and incoherent scattering cross sections of the atomic ensemble. Note that the incoherent scattering cross section is given by $C_{\mathrm{sca}}^{\mathrm{incoh}}=C_{\mathrm{sca}}^{\mathrm{total}}-C_{\mathrm{sca}}^{\mathrm{coh}}$. We consider now a spherical subwavelength atomic cloud with radius $R=0.2\lambda_a$ composed of 25 atoms which can be fully characterized by dipole and quadrupole moments. Figure~\ref{fig:Average_polarizability} (c) shows that the atomic cloud exhibits strong electric and magnetic responses. Figure~\ref{fig:Average_polarizability}(d) shows the coherent, incoherent and total scattering cross sections (normalized to $\lambda^2/2\pi$) calculated from Eq.~(\ref{eq:C_sca_total_1PW_MT}), and the contribution of different multipole moments as a function of frequency detuning. It can be seen that the maximum total scattering cross section of the ensemble is approximately equal to the scattering of a single atom, even though the atomic cloud consists of 25 atoms~\cite{Schilder:2020}. Furthermore, the maximum cross section of coherent scattering is much smaller than that of a single atom~\cite{Schilder:2016,Schilder:2017,Schilder:2020}.

%%%%%%%%%%%%%% Figure 3 %%%%%%%%%%%%%%
\begin{figure*}
    \centering
    \includegraphics[width=0.75\textwidth]{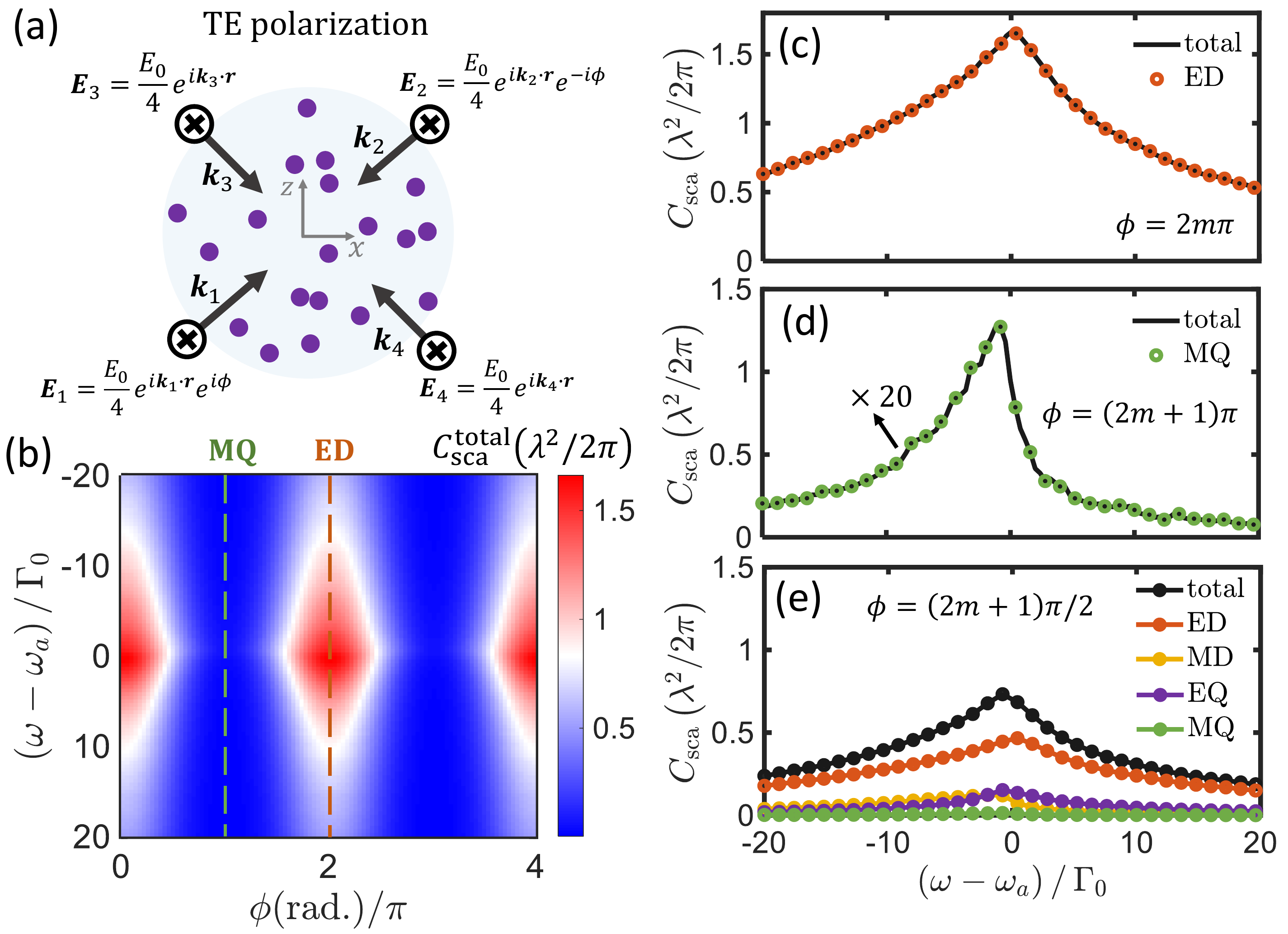}
    \caption{Selective excitation of ED and MQ moments using four TE polarized plane waves. (a) Schematic drawing of a subwavelength atomic cloud when illuminated by four plane waves. $\phi$ is the relative phase between the plane waves. (b) Normalized ensemble-averaged scattering cross section of the atomic cloud as a function of frequency detuning and the relative phase $\phi$. The red and green dashed lines depict selective excitation of pure ED and MQ moments, respectively. (c)-(e) Normalized ensemble-averaged total scattering cross sections as a function of frequency detuning for different phase (c) $\phi=2m\pi$, (d) $\phi=(2m+1)\pi$, and (e) $\phi=(2m+1)\pi/2$. The ensemble-averages are obtained from 10000 realizations of the atomic cloud.}
    \label{fig:3PW_TE}
\end{figure*}
%%%%%%%%%%%%%%%%%%%%%%%%%%%%%%%%%%%%%%

In order to establish the relation between the coherent and incoherent polarizabilities, we focus only on the dipolar response of the atomic cloud for simplicity. The supplementary material provides the relations for other multipole moments. Note that the ensemble-averaged polarizability of a spherical atomic cloud is isotropic, i.e. $\left\langle \bm{\alpha}_{i}\right\rangle =\left\langle \alpha_{i}\right\rangle \mathbf{I}$, where $\mathbf{I}$ is the identity matrix. Therefore, all the diagonal matrix elements of the induced electric dipole polarizability are identical and are represented by $\alpha_{_{\mathrm{ED}}}^{\mathrm{D}}$. All the off-diagonal elements, $\alpha_{_{\mathrm{ED}}}^{\mathrm{OD}}$, are also identical. Therefore, the induced electric dipole polarizability can be written as \begin{equation}
\bm{\alpha}_{_{\mathrm{ED}}}=\left(\left\langle \alpha_{_{\mathrm{ED}}}\right\rangle +\delta\alpha_{_{\mathrm{ED}}}^{\mathrm{D}}\right)\mathbf{I}+\delta\alpha_{_{\mathrm{ED}}}^{\mathrm{OD}}\left(\mathbf{J}-\mathbf{I}\right),\label{polarizability_D_OD}
\end{equation}
where $\mathbf{J}$ is the all-ones matrix. 
Note also that
$\left\langle \bm{\alpha}_{_{\mathrm{ED}}}\right\rangle =\left\langle \alpha_{_{\mathrm{ED}}}\right\rangle \mathbf{I}\,\,$ and $\left\langle \alpha_{_{\mathrm{ED}}}^{\mathrm{OD}}\right\rangle =0$. Using conservation of energy, we get~(see Appendix~\ref{APP_ME})
\begin{eqnarray}
\mathrm{Im}\left(\alpha_{0}\left\langle \alpha_{_{\mathrm{ED}}}\right\rangle \right)-\left|\left\langle \alpha_{_{\mathrm{ED}}}\right\rangle \right|^{2}&=&\left\langle \left|\delta\alpha_{_{\mathrm{ED}}}^{\mathrm{D}}\right|^{2}\right\rangle +2\left\langle \left|\delta\alpha_{_{\mathrm{ED}}}^{\mathrm{OD}}\right|^{2}\right\rangle. \label{eq:OT_ED}
\end{eqnarray}
Equation~(\ref{eq:OT_ED}) is the second main result of this paper; it shows how the fluctuations of the polarizabilities can be obtained from the ensemble-averaged values~(see Appendix \ref{APP_ME}) for the details of the derivation and similar expressions for MD, EQ, MQ polarizability tensors). In Fig.~\ref{fig:polarizability_Coh_incoh}(a)-(b), we plot the coherent and incoherent electric dipole polarizabilities retrieved from multipolar decomposition. In contrast to the off-diagonal terms, the diagonal term exhibits a non-zero ensemble-averaged polarizabiltiy. Note that the components of the electric dipole moment tensor satisfy Eq.~(\ref{eq:OT_ED})~[see Fig.~\ref{fig:polarizability_Coh_incoh}(c)]. Using the duality of the electric and magnetic fields in Maxwell’s equations and conservation of energy, we can obtain a similar relation for the components of the magnetic polarizability tensor~(i.e., replacing MD with ED in Eq.~(\ref{eq:OT_ED}), see Appendix \ref{APP_ME}). Figure~\ref{fig:polarizability_Coh_incoh}(d)-(e) shows the coherent and incoherent components of the magnetic polarizabilities. The magnetic response is smaller than the electric one. The induced multipole moments exhibit asymmetry in their resonance lineshape which explains the non-Lorentzian lineshape of the scattering cross sections in Fig.~\ref{fig:Average_polarizability}(d).
%%%%%%%%%%%%%% Figure 4 %%%%%%%%%%%%%%
\begin{figure*}
    \centering
    \includegraphics[width=0.75\textwidth]{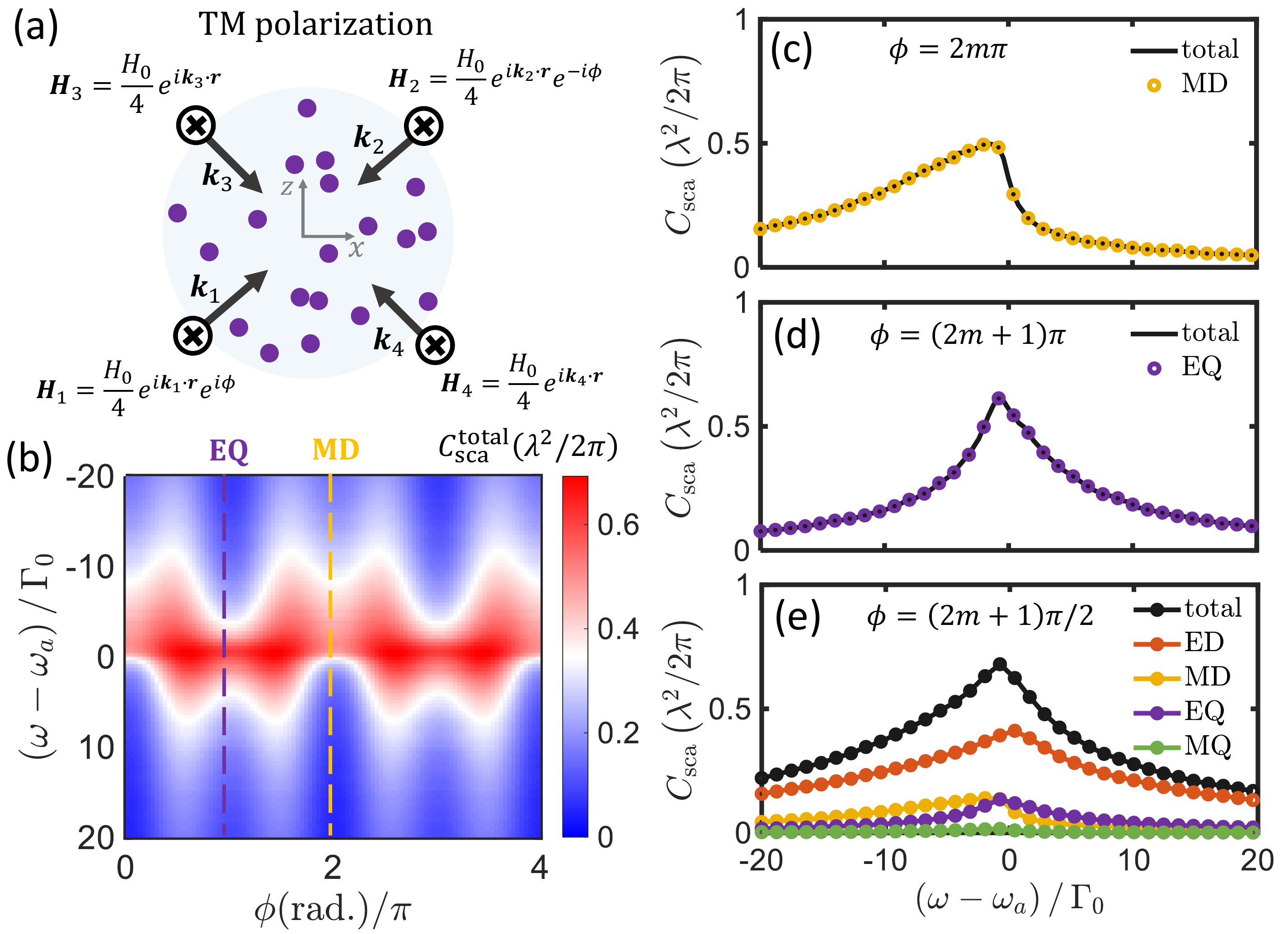}
    \caption{Selective excitation of MD and EQ moments using four TM polarized plane waves. (a) Schematic drawing of a subwavelength atomic cloud when illuminated by four plane waves. $\phi$ is the relative phase between the plane waves. (b) Normalized ensemble-averaged scattering cross section of the atomic cloud as a function of frequency detuning and the relative phase between input plane waves. 
    The orange and purple dashed lines depict selective excitation of pure MD and EQ moments, respectively. (c)-(e) Normalized ensemble-averaged scattering cross sections as a function of frequency detuning for different phase (c) $\phi=2m\pi$, (d) $\phi=(2m+1)\pi$, and (e) $\phi=(2m+1)\phi/2$. The ensemble-averages are obtained from 10000 realizations of the atomic cloud.}
    \label{fig:4PW_TM}
\end{figure*}
%%%%%%%%%%%%%%%%%%%%%%%%%%%%%%%%%%%%%%
\section{Selective excitation of electric dipole or magnetic quadrupole moment}
Although the constituent atoms have only electric dipole transitions, the entire atomic cloud can support higher order electric and magnetic multipole moments~[see Fig.~\ref{fig:Average_polarizability} (d)]. Here, we show that it is possible to selectively excite a particular multipole moment by tailoring the excitation field. To this end, we consider an excitation by four plane waves with TE polarization, i.e., $\mathbf{E}_{\mathrm{inc}}	=	E_{0}/4\sum_{n=1}^{4}e^{i\left(\mathbf{k}_{n}\cdot\mathbf{r}+\phi_{n}\right)}\mathbf{e}_{y},$
where $\mathbf{k}_{1}=-\mathbf{k}_{2}=k\left(\mathrm{sin}\psi\mathbf{e}_{x}+\mathrm{cos}\psi\mathbf{e}_{z}\right)$,  $\mathbf{k}_{3}=-\mathbf{k}_{4}=k\left(\mathrm{sin}\psi\mathbf{e}_{x}-\mathrm{cos}\psi\mathbf{e}_{z}\right),$ $\phi_1=-\phi_2\equiv\phi$, $\phi_3=\phi_4=0$ and $\psi=\pi/4$~[see Fig.~\ref{fig:3PW_TE} (a)]. Hence, the ensemble-averaged induced multipole moments are given by~(see Appendix~\ref{APP_2PW})
\begin{eqnarray}
\left\langle \mathbf{d}^{\mathrm{E}}\right\rangle  & = & \varepsilon_{0}\left\langle \alpha_{_{\mathrm{ED}}}\right\rangle E_{0}\mathrm{cos}^{2}\frac{\phi}{2}\mathbf{e}_{y},\nonumber \\
\left\langle \mathbf{d}^{\mathrm{M}}\right\rangle  & = & -i\left\langle \alpha_{_{\mathrm{MD}}}\right\rangle H_{0}\mathrm{sin}\frac{\phi}{2}\mathrm{cos}\frac{\phi}{2}\left(\mathbf{e}_{x}-\mathbf{e}_{z}\right)/\sqrt{2},\nonumber \\
\left\langle \mathbf{Q}^{\mathrm{E}}\right\rangle  & = & -\varepsilon_{0}\frac{k}{2}\left\langle \alpha_{_{\mathrm{EQ}}}\right\rangle E_{0}\mathrm{sin}\frac{\phi}{2}\mathrm{cos}\frac{\phi}{2}\left\{ \mathbf{e}_{y}\mathbf{e}_{x}+\mathbf{e}_{x}\mathbf{e}_{y}\right.\nonumber \\
 &  & +\left.\mathbf{e}_{y}\mathbf{e}_{z}+\mathbf{e}_{z}\mathbf{e}_{y}\right\} /\sqrt{2},\nonumber \\
\left\langle \mathbf{Q}^{\mathrm{M}}\right\rangle  & = & -\frac{ik}{2}\left\langle \alpha_{_{\mathrm{MQ}}}\right\rangle H_{0}\mathrm{sin}^{2}\frac{\phi}{2}\left(\mathbf{e}_{z}\mathbf{e}_{z}-\mathbf{e}_{x}\mathbf{e}_{x}\right).\label{eq:dQ_2PW}
\end{eqnarray}
Equation (\ref{eq:dQ_2PW}) clearly shows that by changing the relative phase $\phi$, one can control which multipole moment to be excited. Consequently, the scattering cross sections are given by~(see Appendix~\ref{APP_2PW})
\begin{eqnarray}
C_{\mathrm{sca}}^{\mathrm{coh}}&=&\frac{3\lambda^{2}}{2\pi\alpha_{0}^{2}}\left(\left|\left\langle \alpha_{_{\mathrm{ED}}}\right\rangle \right|^{2}\mathrm{cos}^{4}\frac{\phi}{2}+\left|\left\langle \alpha_{_{\mathrm{MD}}}\right\rangle \right|^{2}\mathrm{sin}^{2}\frac{\phi}{2}\mathrm{cos}^{2}\frac{\phi}{2}\right)\nonumber\\
 &  &+\frac{5\lambda^{2}}{2\pi\alpha_{0}^{\prime2}}\left(\left|\left\langle \alpha_{_{\mathrm{EQ}}}\right\rangle \right|^{2}\mathrm{sin}^{2}\frac{\phi}{2}\mathrm{cos}^{2}\frac{\phi}{2}+\left|\left\langle \alpha_{_{\mathrm{MQ}}}\right\rangle \right|^{2}\mathrm{sin}^{4}\frac{\phi}{2}\right),\nonumber\\
C_{\mathrm{sca}}^{\mathrm{total}}&=&\frac{3\lambda^{2}}{2\pi\alpha_{0}}\mathrm{Im}\left(\left\langle \alpha_{_{\mathrm{ED}}}\right\rangle \mathrm{cos}^{4}\frac{\phi}{2}+\left\langle \alpha_{_{\mathrm{MD}}}\right\rangle \mathrm{sin}^{2}\frac{\phi}{2}\mathrm{cos}^{2}\frac{\phi}{2}\right)\label{eq:C_sca_total_4PW_TE_MT}\\
 &  & +\frac{5\lambda^{2}}{2\pi\alpha_{0}^{\prime}}\mathrm{Im}\left(\left\langle \alpha_{_{\mathrm{EQ}}}\right\rangle \mathrm{sin}^{2}\frac{\phi}{2}\mathrm{cos}^{2}\frac{\phi}{2}+\left\langle \alpha_{_{\mathrm{MQ}}}\right\rangle \mathrm{sin}^{4}\frac{\phi}{2}\right).\nonumber
\end{eqnarray}
Equations (\ref{eq:dQ_2PW}) and (\ref{eq:C_sca_total_4PW_TE_MT}) are the third main result of this paper which show that the induced dipole moments and the scattering cross sections can be controlled by a simple four-beam configuration and the relative phase $\phi$ between the plane waves. Figure~\ref{fig:3PW_TE}(b) plots the scattering cross section as a function of the relative phase and the frequency detuning. Interestingly, as can be seen from Fig.~\ref{fig:3PW_TE}(c) and (d), the cooperative resonance linewidth can also be tuned by varying the phase $\phi$ due to selective exciation of different multipole moments.

We note three different scenarios based on the relative phase $\phi$:

(i) At $\phi=2m\pi$, where $m$ is a non-negative integer, only the electric dipole moment of the atomic cloud is excited~[see Eq.~(\ref{eq:dQ_2PW}) and  Fig.~\ref{fig:3PW_TE}(c)]. In this case, the atomic cloud exhibits an omnidirectional radiation pattern. 

(ii) At $\phi=(2m+1)\pi$, according to Eq.~(\ref{eq:dQ_2PW}), the atomic cloud exhibits only a magnetic quadrupole moment as shown in Fig.~\ref{fig:3PW_TE}(d) and scatters light with a quadrupolar pattern. 

(iii) At $2m\pi<\phi<\left(2m+1\right)\pi$, all multipoles can be excited, see for example Fig.~\ref{fig:3PW_TE}(e) for $\phi=(2m+1)\pi/2$. 
Thus, one can selectively excite the electric dipole or magnetic quadrupole moment of the atomic cloud by just controlling the relative phase of the plane waves with TE polarizations and achieve arbitrary radiation patterns.

\section{Selective excitation of magnetic dipole or electric quadrupole moment}To selectively excite magnetic dipole or electric quadrupole, we employ superposition of four plane waves with TM polarization: $\mathbf{H}_{\mathrm{inc}}	=	H_{0}/4\sum_{n=1}^{4}e^{i\left(\mathbf{k}_{n}\cdot\mathbf{r}+\phi_{n}\right)}\mathbf{e}_{y},$ where $\mathbf{k}_n$  and $\phi_n$ are defined similar to the TE polarization, see the previous section and Fig.~\ref{fig:4PW_TM}(a). The coherent and total scattering cross sections are thus given by~(see Appendix~\ref{APP_2PW})

\begin{eqnarray}
C_{\mathrm{sca}}^{\mathrm{coh}} & = & \frac{3\lambda^{2}}{2\pi\alpha_{0}^{2}}\left(\left|\left\langle \alpha_{_{\mathrm{ED}}}\right\rangle \right|^{2}\mathrm{sin}^{2}\frac{\phi}{2}\mathrm{cos}^{2}\frac{\phi}{2}+\left|\left\langle \alpha_{_{\mathrm{MD}}}\right\rangle \right|^{2}\mathrm{cos}^{4}\frac{\phi}{2}\right)\nonumber\\
 &  &+\frac{5\lambda^{2}}{2\pi\alpha_{0}^{\prime2}}\left(\left|\left\langle \alpha_{_{\mathrm{EQ}}}\right\rangle \right|^{2}\mathrm{sin}^{4}\frac{\phi}{2}+\left|\left\langle \alpha_{_{\mathrm{MQ}}}\right\rangle \right|^{2}\mathrm{sin}^{2}\frac{\phi}{2}\mathrm{cos}^{2}\frac{\phi}{2}\right),\nonumber\\
C_{\mathrm{sca}}^{\mathrm{total}} & = & \frac{3\lambda^{2}}{2\pi\alpha_{0}}\mathrm{Im}\left(\left\langle \alpha_{_{\mathrm{ED}}}\right\rangle \mathrm{sin}^{2}\frac{\phi}{2}\mathrm{cos}^{2}\frac{\phi}{2}+\left\langle \alpha_{_{\mathrm{MD}}}\right\rangle \mathrm{cos}^{4}\frac{\phi}{2}\right)\label{eq:C_sca_total_4PW_TM_MT}\\
 &  &+\frac{5\lambda^{2}}{2\pi\alpha_{0}^{\prime}}\mathrm{Im}\left(\left\langle \alpha_{_{\mathrm{EQ}}}\right\rangle \mathrm{sin}^{4}\frac{\phi}{2}+\left\langle \alpha_{_{\mathrm{MQ}}}\right\rangle \mathrm{sin}^{2}\frac{\phi}{2}\mathrm{cos}^{2}\frac{\phi}{2}\right).\nonumber
\end{eqnarray}

Figure~\ref{fig:4PW_TM}(b) plots the total scattering cross section as a function of the relative phase $\phi$ and the frequency detuning. As shown in Fig.~\ref{fig:4PW_TM}(c) and \ref{fig:4PW_TM}(d), by varying the phase $\phi$, one can excite different multipole moments and thus control the cooperative shift and resonance linewidth. In particular, the MD and EQ moments can be selectively excited by the TM polarized plane waves. As a consequence, the atomic cloud will scatter light in a selective direction depending on the relative phase between the plane waves.

Arrays of cold atoms with subwavelength spacing scatter light coherently and thus have been modeled by effective electric and magnetic multipole moments~\cite{Alaee2020,Alaee:2020Kerker,Ballantine2020}. In contrast, an atomic cloud composed of randomly distributed atoms exhibits not only coherent, but also incoherent scattering due to the motion of the atoms~\cite{Schilder:2016,Schilder:2017,Schilder:2020}. In this paper, we showed that the multipolar decomposition can model \textit{not} only the coherent, but also the incoherent response of the atomic cloud accurately. We also demonstrated that the ensemble-averaged polarizabilities are adequate to model the response of the atomic cloud. Furthermore, using superposition of plane waves, we showed that one can selectively excite the induced electric and magnetic multipole moments and thus manipulate the resonant linewidth and cooperative shift of the ensemble, as well as its radiation pattern. Our study paves the way towards controlling cooperative effects in atomic systems through structured light~\cite{Andrews2011Book,Dunlop2016}. Our approach to control the cooperative effects is \textit{not} restricted to subwavelength cold atomic clouds and can be realized both experimentally and theoretically in different systems of interacting quantum emitters including ultracold quantum metasurfaces~\cite{Rui:2020}, nanoscale atomic vapor layer~\cite{Peyrot2018}, two-dimensional semiconductors heterostructures~\cite{Scuri2020}, and atomic arrays in waveguides and cavities~\cite{Sheremet2021waveguide}.

%%%%%%%%%%%%%%%%%%%%%%%%%%%%%%%%%%%%%%

\textbf{Acknowledgments.—} R. A. is grateful to Vahid Sandoghdar, Boris Braverman, and Zeinab Mokhtari for helpful discussions and acknowledges the support of the Alexander von Humboldt Foundation through the Feodor Lynen (Return) Research Fellowship. R.A., A.S., and R.W.B. acknowledge support through the Natural Sciences and Engineering Research Council of Canada, the Canada Research Chairs program, and the Canada First Research Excellence Fund. 

\section{Appendices} 
\appendix
\section{Atomic polarizability and coupled-dipole equations}
Let us consider an atomic cloud composed of neutral atoms with only $\textit{electric}$ dipole transition moments and illuminated by a plane wave~[see Fig.~\ref{fig:Average_polarizability} (a)]. The atoms confined in a volume smaller than the wavelength of the resonant light,
i.e., $D<\lambda$. We consider
the weak-excitation limit where the atomic response is isotropic and
linear. The electric polarizability of each atom amounts to $\alpha\left(\omega\right)=-\left(\alpha_{0}\Gamma_{0}/2\right)/\left[\omega-\omega_{a}+i\left(\Gamma_{0}+\Gamma_{\mathrm{nr}}\right)/2\right]$,
where $\Gamma_{0}$ is the radiative linewidth of the atomic transition at frequency $\omega_{a}$, and $\omega-\omega_{a}\ll\omega_{a}$
represents the frequency detuning between the illumination and the
atomic resonance, $\alpha_{0}=6\pi/k^{3}$ and $k$ is the wavenumber~\cite{lagendijk1996,Lagendijk1998}.
We assume elastic scattering events and therefore the non-radiative
decay rate is zero, i.e., $\Gamma_{\mathrm{nr}}=0$. 
The induced
dipole moment of the $i$th atom $\mathbf{p}\left(\mathbf{r}_{i}\right)=  \epsilon_{0}\alpha\mathbf{E}_\mathrm{loc}\left(\mathbf{r}_{i}\right)$ can be obtained by using the coupled-dipole
equations~\cite{lagendijk1996,Lagendijk1998,Alaee:2017Review}
\begin{eqnarray}
\mathbf{p}\left(\mathbf{r}_{i}\right) & = & \epsilon_{0}\alpha\left[\mathbf{E}_{\mathrm{inc}}\left(\mathbf{r}_{i}\right)+\sum_{i\neq j}\mathbf{G}\left(\mathbf{r}_{i},\mathbf{r}_{j}\right)\mathbf{p}\left(\mathbf{r}_{j}\right)\right],\label{eq:CDT_S}
\end{eqnarray}
where $\mathbf{E}_{\mathrm{inc}}\left(\mathbf{r}_{i}\right)$ is the
incident field at the position $\mathbf{r}_i$ of the atom, and $\alpha$ is the atomic polarizability. The total field at the position of the $i$th atom $\mathbf{E}_\mathrm{loc}\left(\mathbf{r}_{i}\right)$ is the sum of the incident field and the scattered field from the other atoms. The electric dipole at position $\mathbf{r}_{j}$ radiates an electromagnetic field which when measured at $\mathbf{r}_{i}$ can be calculated from $\mathbf{G}\left(\mathbf{r}_{i},\mathbf{r}_{j}\right)\mathbf{p}\left(\mathbf{r}_{j}\right)$, where $\mathbf{G}\left(\mathbf{r}_{i},\mathbf{r}_{j}\right)$ is Green's tensor given by~\cite{tai1994dyadic,Jackson1999}
\begin{equation}
\text{\ensuremath{\mathbf{G}\left(\mathbf{r}_{i},\mathbf{r}_{j}\right)}=\ensuremath{\frac{3}{2\alpha_{0}\epsilon_{0}}e^{i\zeta}\left[g_{1}\left(\zeta\right)\bar{\bar{{\bf I}}}+g_{2}\left(\zeta\right)\mathbf{\mathbf{n}n}\right]}},
\end{equation}
where
\begin{eqnarray}
g_{1}\left(\zeta\right) & = & \left(\frac{1}{\zeta}-\frac{1}{\zeta^{3}}+\frac{i}{\zeta^{2}}\right),\,\,\,\,\,\,\,\,\,\nonumber \\
g_{2}\left(\zeta\right)  &= &  \left(-\frac{1}{\zeta}+\frac{3}{\zeta^{3}}-\frac{3i}{\zeta^{2}}\right),
\end{eqnarray}
$\bar{\bar{{\bf I}}}$ is the identity dyadic, $\mathbf{n}=\frac{\mathbf{r}_{i}-\mathbf{r}_{j}}{\left|\mathbf{r}_{i}-\mathbf{r}_{j}\right|}$,
and $\zeta=\left|k\left(\mathbf{r}_{i}-\mathbf{r}_{j}\right)\right|$~\cite{Alaee2020,Alaee:2020Kerker}. Having the induced dipole moment of each atom, we can define the induced displacement current  $\mathbf{J}\left(\mathbf{r},\omega\right)=-i\omega \sum_{i=1}^{N}\mathbf{p}\left(\mathbf{r}_{i}\right)\delta\left(\mathbf{r}-\mathbf{r}_{i}\right)$, where $\delta$ is the Dirac delta function, and $\mathbf{p}\left(\mathbf{r}_{i}\right)$
is the induced electric dipole moment of the $i$th atom at $\mathbf{r}=\mathbf{r}_{i}$~[see Fig.~\ref{fig:Average_polarizability} (a)]. Here,
we assumed $e^{-i\omega t}$ as a time harmonic variation.
\section{Multipole expansion and cross sections} \label{APP_ME}
\subsection{Coherent and incoherent multipole moments}
In this subsection, we present expressions for the \textit{effective} induced electric and magnetic moments in Cartesian coordinates~\cite{Alaee2020}.
\begin{widetext}
 Using the multipole expansion of the induced current $\mathbf{J}\left(\mathbf{r},\omega\right)$, the induced effective multipole moments of the atomic cloud (at the center $\mathbf{r}=0$) can be calculated~\cite{Alaee:2018,Alaee2019}:

\begin{eqnarray}
d^{\rm E}_{\mu} &=&-\frac{1}{i\omega}\left\{ \int d^{3}\mathbf{r}J_{\mu}j_{0}\left(kr\right)+\frac{k^{2}}{2}\int d^{3}\mathbf{r}\left[3\left(\mathbf{r}\cdot\mathbf{J}\right)r_{\mu}-r^{2}J_{\mu}\right]\frac{j_{2}\left(kr\right)}{\left(kr\right)^{2}}\right\},\nonumber\\
d^{\rm M}_{\mu}&=&\frac{3}{2}\int d^{3}\mathbf{r}\left(\mathbf{r}\times\mathbf{J}\right)_{\mu}\frac{j_{1}\left(kr\right)}{kr},\nonumber\\
Q^{\rm E}_{\mu\nu} & = & -\frac{3}{i\omega}\left\{ \int d^{3}\mathbf{r}\left[3\left(r_{\nu}J_{\mu}+r_{\mu}J_{\nu}\right)-2\left(\mathbf{r}\cdot\mathbf{J}\right)\delta_{\mu\nu}\right]\frac{j_{1}\left(kr\right)}{kr}\right.\nonumber\\
 &  & \left.+2k^{2}\int d^{3}\mathbf{r}\left[5r_{\mu}r_{\nu}\left(\mathbf{r}\cdot\mathbf{J}\right)-\left(r_{\mu}J_{\nu}+r_{\nu}J_{\mu}\right)r^{2}-r^{2}\left(\mathbf{r}\cdot\mathbf{J}\right)\delta_{\mu\nu}\right]\frac{j_{3}\left(kr\right)}{\left(kr\right)^{3}}\right\},\nonumber\\
 Q^{\rm M}_{\mu\nu}&=&15\int d^{3}\mathbf{r}\left\{ r_{\mu}\left(\mathbf{r}\times\mathbf{J}\right)_{\nu}+r_{\nu}\left(\mathbf{r}\times\mathbf{J}\right)_{\mu}\right\} \frac{j_{2}\left(kr\right)}{\left(kr\right)^{2}}, \label{eq:ME_Exact}
\end{eqnarray}
\end{widetext}
where, $\mu,\nu \in {x,y,z}$. The quantities $d^{\rm E}_{\mu}$, $d^{\rm M}_{\mu}$, $Q^{\rm E}_{\mu\nu}$, and $Q^{\rm M}_{\mu\nu}$ are the
electric dipole (ED), magnetic dipole (MD), electric quadrupole (EQ), and magnetic quadrupole (MQ) multipole moments, respectively. $j_{n}$ are the spherical Bessel functions. Note that $\mathbf{Q}^{\mathrm{E}}=\sum_{\mu,\nu} Q^{\rm E}_{\mu\nu}\mathbf{e}_{\mu}\mathbf{e}_{\nu}$ and $\mathbf{Q}^{\mathrm{M}}=\sum_{\mu,\nu} Q^{\rm M}_{\mu\nu}\mathbf{e}_{\mu}\mathbf{e}_{\nu}$ are tensors of rank two and $\mathbf{e}_{\mu}\mathbf{e}_{\nu}$ is unit dyad. We consider $N_{\rm R}$ realizations for which the positions of the atoms are changed with a uniform probability distribution in a spherical volume. Then, the induced multipole moments of the atomic cloud can be decomposed into coherent (ensemble-averaged) and incoherent (fluctuating) parts:
\begin{eqnarray}
d^{\rm E}_{\mu} & = & \left\langle d^{\rm E}_{\mu}\right\rangle +\delta d^{\rm E}_{\mu},\,\,\,\,\,\,\,\,\,\,d^{\rm M}_{\mu}=\left\langle d^{\rm M}_{\mu}\right\rangle +\delta d^{\rm M}_{i},\label{eq:ME_coh_incoh} \\
Q^{\rm E}_{\mu\nu} & = & \left\langle Q^{\rm E}_{\mu\nu}\right\rangle +\delta Q^{\rm E}_{\mu\nu},\,\,\,\,\,\,\,\,\,\,Q^{\rm M}_{\mu\nu}=\left\langle Q^{\rm M}_{\mu\nu}\right\rangle +\delta Q^{\rm M}_{\mu\nu},\nonumber 
\end{eqnarray}
where $\mu,\nu=x,y,z$. The symbols $\left\langle \cdot\right\rangle $ represent the ensemble-averaged multipole moments. Note that the incoherent multipole moments are related to the quasi-isotropic speckle originating from
the random positions of the atoms in the spherical cloud. 

%%The standard deviations for each multipole moments are given by $\sqrt{\left\langle \left|\delta d_{\mu}^{\mathrm{E}}\right|^{2}\right\rangle}$, $\sqrt{\left\langle \left|\delta d_{\mu}^{\mathrm{M}}\right|^{2}\right\rangle}$, $\sqrt{\left\langle \left|\delta Q_{\mu\nu}^{\mathrm{E}}\right|^{2}\right\rangle }$, and  $\sqrt{\left\langle \left|\delta Q_{\mu\nu}^{\mathrm{M}}\right|^{2}\right\rangle }$.

%%%%%%%%%%%%%%%%%%%%%%%%%%%%%%%%%%%%%%%%
%%%%%%%%%%% New section %%%%%%%%%%%%%%%%%
\subsection{Coherent and incoherent cross sections}

In this subsection, we derive coherent and incoherent scattering and
extinction cross sections using the induced electric and magnetic
multipole moments in Eqs.~(\ref{eq:ME_Exact})-(\ref{eq:ME_coh_incoh}). The total scattering cross section can be decomposed into coherent
and incoherent parts, i.e., $C_{\mathrm{sca}}^{\mathrm{total}}=C_{\mathrm{sca}}^{\mathrm{coh}}+C_{\mathrm{sca}}^{\mathrm{incoh}}$ which are and given by~\cite{Alaee:2018,Alaee2019}
\begin{eqnarray}
C_{\mathrm{sca}}^{\mathrm{coh}}	& = &	\frac{k^{4}}{6\pi\varepsilon_{0}^{2}E_{0}^{2}}\sum_{\mu}\left(\left|\left\langle d_{\mu}^{\mathrm{E}}\right\rangle \right|^{2}+\left|\left\langle \frac{d_{\mu}^{\mathrm{M}}}{c}\right\rangle \right|^{2}\right)\label{eq:C_sca_ME_coh}\\
& & +\frac{k^{6}}{720\pi\varepsilon_{0}^{2}E_{0}^{2}}\sum_{\mu,\nu}\left(\left|\left\langle Q_{\mu\nu}^{\mathrm{E}}\right\rangle \right|^{2}+\left|\left\langle \frac{Q_{\mu\nu}^{\mathrm{M}}}{c}\right\rangle \right|^{2}\right),\nonumber\\
C_{\mathrm{sca}}^{\mathrm{incoh}} & = &	\frac{k^{4}}{6\pi\varepsilon_{0}^{2}E_{0}^{2}}\sum_{\mu}\left(\left\langle \left|\delta d_{\mu}^{\mathrm{E}}\right|^{2}\right\rangle +\left\langle \left|\frac{\delta d_{\mu}^{\mathrm{M}}}{c}\right|^{2}\right\rangle \right)\label{eq:C_sca_ME_incoh}\\
&  &+\frac{k^{6}}{720\pi\varepsilon_{0}^{2}E_{0}^{2}}\sum_{\mu,\nu}\left(\left\langle \left|\delta Q_{\mu\nu}^{\mathrm{E}}\right|^{2}\right\rangle +\left\langle \left|\frac{\delta Q_{\mu\nu}^{\mathrm{M}}}{c}\right|^{2}\right\rangle \right),\nonumber
\end{eqnarray}
and the extinction cross section of the cloud is given by~\cite{Alaee:2018,Alaee2019}
\begin{eqnarray}
C_{\mathrm{ext}} & = &	\frac{k}{\varepsilon_{0}E_{0}^{2}}\mathrm{Im}\left[\sum_{\mu}\left(\left\langle d_{\mu}^{\mathrm{E}}\right\rangle E_{\mu}^{*}+\left\langle \frac{d_{\mu}^{\mathrm{M}}}{c}\right\rangle Z_{0}H_{\mu}^{*}\right)\right]\label{eq:C_ext_ME_total}\\
& &	+\frac{k}{12\varepsilon_{0}E_{0}^{2}}\mathrm{Im}\left[\sum_{\mu,\nu}\left\langle Q_{\mu\nu}^{\mathrm{E}}\right\rangle \left(\frac{\partial E_{\nu}^{*}}{\partial r_{\mu}}+\frac{\partial E_{\mu}^{*}}{\partial r_{\nu}}\right)\right]\nonumber\\
& &+\frac{k}{12\varepsilon_{0}E_{0}^{2}}\mathrm{Im}\left[\sum_{\mu,\nu}\left\langle \frac{Q_{\mu\nu}^{\mathrm{M}}}{c}\right\rangle Z_{0}\left(\frac{\partial H_{\nu}^{*}}{\partial r_{\mu}}+\frac{\partial H_{\mu}^{*}}{\partial r_{\nu}}\right)\right],\nonumber
\end{eqnarray}
where $Z_0$ is the impedance of free space, $c$ is the speed of light in free space, and $\mathbf{r}=\sum_{\mu}r_{\mu} \mathbf{e}_{\mu}=x\mathbf{e}_{x}+y\mathbf{e}_{y}+z\mathbf{e}_{z}$. Note that in Eq.~(\ref{eq:C_ext_ME_total}) and also in the remainder of the supplementary material, $\mathbf{E}$ and $\mathbf{H}$ show the incident fields, i.e. we omit the subscript "inc" for simplifying the notation.

\subsection{Conservation of energy: coherent and incoherent cross sections}
According to the conservation of energy, the extinction cross section is equal to the sum of the coherent and incoherent scattering cross sections, i.e., $C_{\mathrm{ext}}=C_{\mathrm{sca}}^{\mathrm{total}}=C_{\mathrm{sca}}^{\mathrm{coh}}+C_{\mathrm{sca}}^{\mathrm{incoh}}$. Therefore, from Eqs.~(\ref{eq:C_sca_ME_coh})-(\ref{eq:C_ext_ME_total}), we obtain the following relations between the coherent and incoherent multipole moments:
\begin{eqnarray}
\sum_{\mu}\left\langle \left|d_{\mu}^{\mathrm{E}}\right|^{2}\right\rangle  & = & \sum_{\mu}\left[\left|\left\langle d_{\mu}^{\mathrm{E}}\right\rangle \right|^{2}+\left\langle \left|\delta d_{\mu}^{\mathrm{E}}\right|^{2}\right\rangle \right]\nonumber\\
& =&\varepsilon_{0}\alpha_{0}\sum_{\mu}\mathrm{Im}\left[\left\langle d_{\mu}^{\mathrm{E}}\right\rangle E_{\mu}^*\right],\nonumber \\
\sum_{\mu}\left\langle \left|d_{\mu}^{\mathrm{M}}\right|^{2}\right\rangle  & = & \sum_{\mu}\left[\left|\left\langle d_{\mu}^{\mathrm{M}}\right\rangle \right|^{2}+\left\langle \left|\delta d_{\mu}^{\mathrm{M}}\right|^{2}\right\rangle \right]\nonumber\\
& =&\alpha_{0}\sum_{\mu}\mathrm{Im}\left[\left\langle d_{\mu}^{\mathrm{M}}\right\rangle H_{\mu}^*\right],\nonumber \\
\sum_{\mu,\nu}\left\langle \left|Q_{\mu\nu}^{\mathrm{E}}\right|^{2}\right\rangle  & = & \sum_{\mu,\nu}\left[\left|\left\langle Q_{\mu\nu}^{\mathrm{E}}\right\rangle \right|^{2}+\left\langle \left|\delta Q_{\mu\nu}^{\mathrm{E}}\right|^{2}\right\rangle \right]\nonumber\\
& =&\frac{1}{2}\varepsilon_{0}\alpha_{0}^{\prime}\sum_{\mu,\nu}\mathrm{Im}\left[\left\langle Q_{\mu\nu}^{\mathrm{E}}\right\rangle \left(\frac{\partial E_{\nu}^{*}}{\partial x_{\mu}}+\frac{\partial E_{\mu}^{*}}{\partial x_{\nu}}\right)\right],\nonumber \\
\sum_{\mu,\nu}\left\langle \left|Q_{\mu\nu}^{\mathrm{M}}\right|^{2}\right\rangle  & = & \sum_{\mu,\nu}\left[\left|\left\langle Q_{\mu\nu}^{\mathrm{M}}\right\rangle \right|^{2}+\left\langle \left|\delta Q_{\mu\nu}^{\mathrm{M}}\right|^{2}\right\rangle \right]\label{eq:ConEn}\\
& =&\frac{1}{2}\alpha_{0}^{\prime}\sum_{\mu,\nu}\mathrm{Im}\left[\left\langle Q_{\mu\nu}^{\mathrm{M}}\right\rangle \left(\frac{\partial H_{\nu}^{*}}{\partial x_{\mu}}+\frac{\partial H_{\mu}^{*}}{\partial x_{\nu}}\right)\right],\nonumber
\end{eqnarray}
where $\alpha_0 = 6\pi/k^3$ ($\alpha_0^{\prime} = 120\pi/k^5$) is related to the radiation loss of a dipole (quadrupole) moment.

\subsection{Coherent and incoherent dipole polarizabilities}
From Eq.~(\ref{eq:ConEn}) the relation between the coherent and incoherent electric dipole polarizabilities is found to be:
\begin{eqnarray}
\left[\left\langle \left|\delta d_{\mu}^{\mathrm{E}}\right|^{2}\right\rangle \right] & = & \varepsilon_{0}\alpha_{0}\mathrm{Im}\left[\left\langle d_{\mu}^{\mathrm{E}}\right\rangle \left\langle E_{\mu}^{*}\right\rangle \right]-\left|\left\langle d_{\mu}^{\mathrm{E}}\right\rangle \right|^{2}.\label{eq:ConEn_ED}
\end{eqnarray}
For a spherical cloud, the averaged electric polarizability tensor
is isotropic and reads as $\left\langle \bm{\alpha}_{\mathrm{_{ED}}}\right\rangle =\left\langle \alpha_{_{\mathrm{ED}}}\right\rangle \mathbf{I}$,
where $\mathbf{I}$ is the identity matrix. Now by substituting $\left\langle d_{\mu}^{\mathrm{E}}\right\rangle =\varepsilon_{0}\left\langle \alpha_{_{\mathrm{ED}}}\right\rangle E_{\mu}$
and $\left|\left\langle d_{\mu}^{\mathrm{E}}\right\rangle \right|^{2}=\varepsilon_{0}^{2}\left|\left\langle \alpha_{_{\mathrm{ED}}}\right\rangle \right|^{2}\left|E_{\mu}\right|^{2}$
into Eq.~(\ref{eq:ConEn_ED}), we get
\begin{eqnarray}
\sum_{\mu,\nu}\left\langle \left|\frac{\delta\alpha_{\mu\nu}^{\mathrm{ED}}}{\alpha_{0}}\right|^{2}\right\rangle \left|\left\langle E_{\nu}\right\rangle \right|^{2}&=&	\mathrm{Im}\left[\left\langle \frac{\alpha_{_{\mathrm{ED}}}}{\alpha_{0}}\right\rangle \right]\sum_{\nu}\left|\left\langle E_{\nu}\right\rangle \right|^{2}\nonumber\\
& &-\left|\left\langle \frac{\alpha_{_{\mathrm{ED}}}}{\alpha_{0}}\right\rangle \right|^{2}\sum_{\nu}\left|\left\langle E_{\nu}\right\rangle \right|^{2},
\end{eqnarray}
For an $x$-polarized plane wave excitation $\mathbf{E}=E_{0}e^{ikz}\mathbf{e}_{x}$,
we get
\begin{eqnarray}
\mathrm{Im}\left[\left\langle \frac{\alpha_{_{\mathrm{ED}}}}{\alpha_{0}}\right\rangle \right]-\left|\left\langle \frac{\alpha_{_{\mathrm{ED}}}}{\alpha_{0}}\right\rangle \right|^{2}& = & \left\langle \left|\frac{\delta\alpha_{xx}^{\mathrm{ED}}}{\alpha_{0}}\right|^{2}\right\rangle +\left\langle \left|\frac{\delta\alpha_{yx}^{\mathrm{ED}}}{\alpha_{0}}\right|^{2}\right\rangle \nonumber\\
& & +\left\langle \left|\frac{\delta\alpha_{zx}^{\mathrm{ED}}}{\alpha_{0}}\right|^{2}\right\rangle .\label{eq:ED_alpha_v1}
\end{eqnarray}
Using the symmetry of the electric dipole polarizability tensor $\delta\alpha_{zx}^{\mathrm{ED}}=\delta\alpha_{yx}^{\mathrm{ED}}=\delta\alpha_{yz}^{\mathrm{ED}}$
and $\left\langle \bm{\alpha}_{\mathrm{ED}}\right\rangle =\left\langle \alpha_{_{\mathrm{ED}}}\right\rangle \mathbf{I}$,
the electric dipole polarizability tensor can be written as
\begin{eqnarray}
\bm{\alpha}_{\mathrm{_{ED}}} & = & \left[\begin{array}{ccc}
\alpha_{xx}^{\mathrm{^{ED}}} & \alpha_{xy}^{\mathrm{^{ED}}} & \alpha_{xz}^{\mathrm{^{ED}}}\\
\alpha_{yx}^{\mathrm{^{ED}}} & \alpha_{yy}^{\mathrm{^{ED}}} & \alpha_{yz}^{\mathrm{^{ED}}}\\
\alpha_{zx}^{\mathrm{^{ED}}} & \alpha_{zy}^{\mathrm{^{ED}}} & \alpha_{zz}^{\mathrm{^{ED}}}
\end{array}\right],\nonumber\\
&=&\left[\begin{array}{ccc}
\left\langle \alpha_{_{\mathrm{ED}}}\right\rangle +\delta\alpha_{_{\mathrm{ED}}}^{\mathrm{D}} & \delta\alpha_{\mathrm{_{ED}}}^{\mathrm{OD}} & \delta\alpha_{\mathrm{_{ED}}}^{\mathrm{OD}}\\
\delta\alpha_{\mathrm{_{ED}}}^{\mathrm{OD}} & \left\langle \alpha_{_{\mathrm{ED}}}\right\rangle +\delta\alpha_{_{\mathrm{ED}}}^{\mathrm{D}} & \delta\alpha_{\mathrm{_{ED}}}^{\mathrm{OD}}\\
\delta\alpha_{\mathrm{_{ED}}}^{\mathrm{OD}} & \delta\alpha_{\mathrm{_{ED}}}^{\mathrm{OD}} & \left\langle \alpha_{_{\mathrm{ED}}}\right\rangle +\delta\alpha_{_{\mathrm{ED}}}^{\mathrm{D}}
\end{array}\right],\nonumber \\
 & = & \alpha_{_{\mathrm{ED}}}^{\mathrm{D}}\mathbf{I}+\alpha_{_{\mathrm{ED}}}^{\mathrm{OD}}\left(\mathbf{J-\mathbf{I}}\right),
\end{eqnarray}
where $\left\langle \bm{\alpha}_{_{\mathrm{ED}}}\right\rangle =\left\langle \alpha_{_{\mathrm{ED}}}\right\rangle \mathbf{I}$ and
$\left\langle \alpha_{_{\mathrm{ED}}}^{\mathrm{OD}}\right\rangle =0$. 
${\bf I}$ and ${\bf J}$ are the identity and all-ones matrices, respectively.
Therefore, Eq.~(\ref{eq:ED_alpha_v1}) can be simplified
as 
\begin{eqnarray}
\mathrm{Im}\left(\alpha_{0}\left\langle \alpha_{_{\mathrm{ED}}}\right\rangle \right)-\left|\left\langle \alpha_{_{\mathrm{ED}}}\right\rangle \right|^{2}&=&\left\langle \left|\delta\alpha_{_{\mathrm{ED}}}^{\mathrm{D}}\right|^{2}\right\rangle +2\left\langle \left|\delta\alpha_{_{\mathrm{ED}}}^{\mathrm{OD}}\right|^{2}\right\rangle,\nonumber
\end{eqnarray}
and we obtain Eq.~(\ref{eq:OT_ED}) of the main text.
Using the duality in Maxwell\textquoteright s equations, a similar expression for a magnetic polarizability can be obtained:
\begin{eqnarray}
\mathrm{Im}\left[\left\langle \frac{\alpha_{_{\mathrm{MD}}}}{\alpha_{0}}\right\rangle \right]-\left|\left\langle \frac{\alpha_{_{\mathrm{MD}}}}{\alpha_{0}}\right\rangle \right|^{2}& = & \left\langle \left|\frac{\delta\alpha_{xy}^{\mathrm{MD}}}{\alpha_{0}}\right|^{2}\right\rangle +\left\langle \left|\frac{\delta\alpha_{yy}^{\mathrm{MD}}}{\alpha_{0}}\right|^{2}\right\rangle \nonumber\\
& & +\left\langle \left|\frac{\delta\alpha_{zy}^{\mathrm{MD}}}{\alpha_{0}}\right|^{2}\right\rangle .\label{eq:MD_alpha_v1}
\end{eqnarray}

\subsection{Coherent and incoherent quadrapole polarizabilities}
In this subsection, we obtain the relation between the
coherent and incoherent quadrupole polarizabilities using Eq.~(\ref{eq:ConEn}):
\begin{eqnarray}
\left\langle \left|\delta Q_{\mu\nu}^{\mathrm{E}}\right|^{2}\right\rangle  & = & \frac{1}{2}\varepsilon_{0}\alpha_{0}^{\prime}\mathrm{Im}\left[\left\langle Q_{\mu\nu}^{\mathrm{E}}\right\rangle \left(\frac{\partial E_{\nu}^{*}}{\partial x_{\mu}}+\frac{\partial E_{\mu}^{*}}{\partial x_{\nu}}\right)\right]\nonumber\\
& &-\left|\left\langle Q_{\mu\nu}^{\mathrm{E}}\right\rangle \right|^{2},\label{eq:ConEn_EQ}
\end{eqnarray}
and by substituting $\left\langle Q_{\mu\nu}^{\mathrm{E}}\right\rangle =\frac{1}{2}\varepsilon_{0}\left\langle \alpha_{_{\mathrm{EQ}}}\right\rangle \left\langle \frac{\partial E_{\nu}}{\partial x_{\mu}}+\frac{\partial E_{\mu}}{\partial x_{\nu}}\right\rangle $
and $\left|\left\langle Q_{\mu\nu}^{\mathrm{E}}\right\rangle \right|^{2}=\frac{1}{4}\varepsilon_{0}^{2}\left|\left\langle \alpha_{_{\mathrm{EQ}}}\right\rangle \right|^{2}\left|\frac{\partial E_{\nu}}{\partial x_{\mu}}+\frac{\partial E_{\mu}}{\partial x_{\nu}}\right|^{2}$
into Eq.~(\ref{eq:ConEn_EQ}), we get
\begin{eqnarray}
\left\langle \left|\delta\alpha_{\mu\nu\beta\gamma}^{\mathrm{EQ}}\right|^{2}\right\rangle \left|\frac{\partial E_{\beta}}{\partial x_{\gamma}}+\frac{\partial E_{\gamma}}{\partial x_{\beta}}\right|^{2}&=&\alpha_{0}^{\prime}\mathrm{Im}\left[\left\langle \alpha_{_{\mathrm{EQ}}}\right\rangle \right]\left|\frac{\partial E_{\nu}}{\partial x_{\mu}}+\frac{\partial E_{\mu}}{\partial x_{\nu}}\right|^{2}\nonumber\\
&& -\left|\left\langle \alpha_{_{\mathrm{EQ}}}\right\rangle \right|^{2}\left|\frac{\partial E_{\nu}}{\partial x_{\mu}}+\frac{\partial E_{\mu}}{\partial x_{\nu}}\right|^{2}.\nonumber
\end{eqnarray}
The electric quadrupole tensor is a tensor and is  given by
\begin{eqnarray}
\bm{Q}^{\mathrm{E}} & = & \left(\begin{array}{ccc}
Q_{xx}^{\mathrm{E}} & Q_{xy}^{\mathrm{E}} & Q_{xz}^{\mathrm{E}}\\
Q_{yx}^{\mathrm{E}} & Q_{yy}^{\mathrm{E}} & Q_{yz}^{\mathrm{E}}\\
Q_{zx}^{\mathrm{E}} & Q_{zy}^{\mathrm{E}} & Q_{zz}^{\mathrm{E}}
\end{array}\right).
\end{eqnarray}
The quadrupole tensor $\bm{Q}^{\mathrm{E}}$ is symmetric,
i.e. $Q_{xy}^{\mathrm{E}}=Q_{yx}^{\mathrm{E}}$, $Q_{xz}^{\mathrm{E}}=Q_{zx}^{\mathrm{E}}$,
$Q_{yz}^{\mathrm{E}}=Q_{zy}^{\mathrm{E}}$ and traceless $Q_{xx}^{\mathrm{E}}+Q_{yy}^{\mathrm{E}}+Q_{zz}^{\mathrm{E}}=0$.
Therefore, $\bm{Q}^{\mathrm{E}}$ has five
independent components in Cartesian coordinates. These five
independent components are represented by $Q_{xx}^{\mathrm{E}},\, Q_{xy}^{\mathrm{E}},\, Q_{xz}^{\mathrm{E}},\, Q_{yy}^{\mathrm{E}},\, Q_{yz}^{\mathrm{E}}$. 
Now, for a single plane wave excitation $\mathbf{E}=E_{0}e^{ikz}\mathbf{e}_{x}$, we have $\nabla\mathbf{E}+\mathbf{E}\nabla=ikE_{0}\left(\mathbf{e}_{x}\mathbf{e}_{z}+\mathbf{e}_{z}\mathbf{e}_{x}\right)$, and we get
\begin{equation}
\sum_{\mu,\nu}\left\langle \left|\frac{\delta\alpha_{\mu\nu xz}^{\mathrm{EQ}}}{\alpha_{0}^{\prime}}\right|^{2}\right\rangle =\mathrm{Im}\left[\left\langle \frac{\alpha_{_{\mathrm{EQ}}}}{\alpha_{0}^{\prime}}\right\rangle \right]-\left|\left\langle \frac{\alpha_{_{\mathrm{EQ}}}}{\alpha_{0}^{\prime}}\right\rangle \right|^{2}.
\end{equation}

Using the duality in Maxwell\textquoteright s equations, a
similar expression can be found for coherent and incoherent magnetic polarizabilities:
\begin{equation}
\sum_{\mu,\nu}\left\langle \left|\frac{\delta\alpha_{\mu\nu xz}^{\mathrm{MQ}}}{\alpha_{0}^{\prime}}\right|^{2}\right\rangle =\mathrm{Im}\left[\left\langle \frac{\alpha_{_{\mathrm{MQ}}}}{\alpha_{0}^{\prime}}\right\rangle \right]-\left|\left\langle \frac{\alpha_{_{\mathrm{MQ}}}}{\alpha_{0}^{\prime}}\right\rangle \right|^{2}.
\end{equation}

\section{ Single plane wave illumination} \label{APP_1PW}
In this section, we provide analytical expressions for
coherent and incoherent scattering cross sections of an atomic cloud when illuminated by a single plane wave.
\subsection{Ensemble-averaged multipole moments}
Let us consider a cloud illuminated by a plane wave
$\mathbf{E}=E_{0}e^{ikz}\mathbf{e}_{x}$ propagating
in the $z$ direction, where $\mathbf{e}_{x}$ is the unit vector
in the $x$ direction. The ensemble-averaged induced multipole
moments of the cloud at $\mathbf{r}=0$ are given by
\begin{eqnarray}
\left\langle \mathbf{d}^{\mathrm{E}}\right\rangle  & = & \varepsilon_{0}\left\langle \alpha_{_{\mathrm{ED}}}\right\rangle \mathbf{E}\left(\mathbf{r}=0\right)=\varepsilon_{0}\left\langle \alpha_{_{\mathrm{ED}}}\right\rangle E_{0}\mathbf{e}_{x},\nonumber \\
\left\langle \mathbf{d}^{\mathrm{M}}\right\rangle  & = & \left\langle \alpha_{_{\mathrm{MD}}}\right\rangle \mathbf{H}\left(\mathbf{r}=0\right)=\left\langle \alpha_{_{\mathrm{MD}}}\right\rangle H_{0}\mathbf{e}_{y},\nonumber \\
\left\langle \mathbf{Q}^{\mathrm{E}}\right\rangle  & = & \frac{1}{2}\varepsilon_{0}\left\langle \alpha_{_{\mathrm{EQ}}}\right\rangle \left.\left(\mathbf{\nabla\mathbf{E}}+\mathbf{\mathbf{E}\nabla}\right)\right|_{\mathbf{r}=0}\nonumber \\
 &=&\left\langle Q_{xz}^{\mathrm{E}}\right\rangle \mathbf{e}_{x}\mathbf{e}_{z}+\left\langle Q_{zx}^{\mathrm{E}}\right\rangle \mathbf{e}_{z}\mathbf{e}_{x}\nonumber \\
 & = & \frac{1}{2}\varepsilon_{0}\left\langle \alpha_{_{\mathrm{EQ}}}\right\rangle \left.\left(\frac{\partial E_{x}}{\partial z}+\frac{\partial E_{z}}{\partial x}\right)\right|_{\mathbf{r}=0}\left(\mathbf{e}_{x}\mathbf{e}_{z}+\mathbf{e}_{z}\mathbf{e}_{x}\right)\nonumber \\
 & =&\frac{ik}{2}\varepsilon_{0}\left\langle \alpha_{_{\mathrm{EQ}}}\right\rangle E_{0}\left(\mathbf{e}_{x}\mathbf{e}_{z}+\mathbf{e}_{z}\mathbf{e}_{x}\right),\nonumber \\
\left\langle \mathbf{Q}^{\mathrm{M}}\right\rangle  & = & \frac{1}{2}\left\langle \alpha_{_{\mathrm{MQ}}}\right\rangle \left.\left(\mathbf{\nabla H}+\mathbf{\mathbf{H}\nabla}\right)\right|_{\mathbf{r}=0}\nonumber\\
&=&\left\langle Q_{yz}^{\mathrm{M}}\right\rangle \mathbf{e}_{y}\mathbf{e}_{z}+\left\langle Q_{zy}^{\mathrm{M}}\right\rangle \mathbf{e}_{z}\mathbf{e}_{y}\nonumber \\
 & = & \frac{1}{2}\left\langle \alpha_{_{\mathrm{MQ}}}\right\rangle \left.\left(\frac{\partial H_{y}}{\partial z}+\frac{\partial H_{z}}{\partial y}\right)\right|_{\mathbf{r}=0}\left(\mathbf{e}_{y}\mathbf{e}_{z}+\mathbf{e}_{z}\mathbf{e}_{y}\right)\nonumber\\&=&\frac{ik}{2}\left\langle \alpha_{_{\mathrm{MQ}}}\right\rangle H_{0}\left(\mathbf{e}_{y}\mathbf{e}_{z}+\mathbf{e}_{z}\mathbf{e}_{y}\right),\label{eq:ME_ave_SPW}
\end{eqnarray}
where $\left\langle \alpha_{_{\mathrm{ED}}}\right\rangle $ ($\left\langle \alpha_{_{\mathrm{MD}}}\right\rangle $)
and $\left\langle \alpha_{_{\mathrm{EQ}}}\right\rangle $ ($\left\langle \alpha_{_{\mathrm{MQ}}}\right\rangle $)
are ensemble-averaged electric (magnetic) dipole and quadrupole polarizabilities,
respectively. $\mathbf{E}$ and $\mathbf{H}$ in Eq.~(\ref{eq:ME_ave_SPW}) are the incident electric and magnetic fields, respectively. 

\subsection{Coherent and incoherent cross sections}
In this subsection, we find the scattering cross sections as a function of ensemble-averaged dipole and quadrupole polarizabilities. By substituting Eq.~(\ref{eq:ME_ave_SPW}) into Eqs.~(\ref{eq:C_sca_ME_coh}) and (\ref{eq:C_ext_ME_total})
we obtain
\begin{eqnarray}
C_{\mathrm{sca}}^{\mathrm{coh}} & = & \frac{k^{4}}{6\pi}\left(\left|\left\langle \alpha_{_{\mathrm{ED}}}\right\rangle \right|^{2}+\left|\left\langle \alpha_{_{\mathrm{MD}}}\right\rangle \right|^{2}\right)\nonumber\\
&&+\frac{k^{8}}{1440}\left(\left|\left\langle \alpha_{_{\mathrm{EQ}}}\right\rangle \right|^{2}+\left|\left\langle \alpha_{_{\mathrm{MQ}}}\right\rangle \right|^{2}\right),\\
C_{\mathrm{ext}} & = & k\mathrm{Im}\left[\left\langle \alpha_{_{\mathrm{ED}}}\right\rangle +\left\langle \alpha_{_{\mathrm{MD}}}\right\rangle+\frac{k^{2}}{12}\left(\left\langle \alpha_{_{\mathrm{EQ}}}\right\rangle +\left\langle \alpha_{_{\mathrm{MQ}}}\right\rangle \right)\right].\nonumber
\end{eqnarray}

After applying some simple algebra and using $\alpha_0=6\pi/k^3$ and $\alpha_0^{\prime}=120\pi/k^5$, we obtain Eq.~(\ref{eq:C_sca_total_1PW_MT}) of the main text:
\begin{eqnarray}
C_{\mathrm{sca}}^{\mathrm{coh}} & = & \frac{3\lambda^{2}}{2\pi}\left(\left|\left\langle \frac{\alpha_{_{\mathrm{ED}}}}{\alpha_{0}}\right\rangle \right|^{2}+\left|\left\langle \frac{\alpha_{_{\mathrm{MD}}}}{\alpha_{0}}\right\rangle \right|^{2}\right)\nonumber\\
&& +\frac{5\lambda^{2}}{2\pi}\left(\left|\left\langle \frac{\alpha_{_{\mathrm{EQ}}}}{\alpha_{0}^{\prime}}\right\rangle \right|^{2}+\left|\left\langle \frac{\alpha_{_{\mathrm{MQ}}}}{\alpha_{0}^{\prime}}\right\rangle \right|^{2}\right),\\\nonumber
C_{\mathrm{ext}} & = & \frac{3\lambda^{2}}{2\pi}\mathrm{Im}\left[\left\langle \frac{\alpha_{_{\mathrm{ED}}}}{\alpha_{0}}\right\rangle +\left\langle \frac{\alpha_{_{\mathrm{MD}}}}{\alpha_{0}}\right\rangle \right]\nonumber\\
&& +\frac{5\lambda^{2}}{2\pi}\mathrm{Im}\left[\left\langle \frac{\alpha_{_{\mathrm{EQ}}}}{\alpha_{0}^{\prime}}\right\rangle +\left\langle \frac{\alpha_{_{\mathrm{MQ}}}}{\alpha_{0}^{\prime}}\right\rangle \right].\label{eq:C_sca_1PW_norm_final}
\end{eqnarray}
Using above equations, we can calculate incoherent
scattering cross section from $C_{\mathrm{sca}}^{\mathrm{incoh}}=C_{\mathrm{ext}}-C_{\mathrm{sca}}^{\mathrm{coh}}$.

\section{ Selective excitation} \label{APP_2PW}

\subsection{Four plane waves with TM polarization} 
In this subsection, we consider an atomic cloud when illuminated by
four plane waves with TM polarization. The magnetic fields of the
plane waves are given by
\begin{eqnarray}
\mathbf{H}_{1} & = & \frac{H_{0}}{4}e^{i\left(\mathbf{k}_{1}\cdot\mathbf{r}+\phi\right)}=\frac{H_{0}}{4}e^{i\left(k_{x}x+k_{z}z+\phi\right)}\mathbf{e}_{y},\nonumber\\\mathbf{H}_{2}& = &\frac{H_{0}}{4}e^{i\left(\mathbf{k}_{2}\cdot\mathbf{r}-\phi\right)}=\frac{H_{0}}{4}e^{-i\left(k_{x}x+k_{z}z+\phi\right)}\mathbf{e}_{y},\nonumber \\
\mathbf{H}_{3} & = & \frac{H_{0}}{4}e^{i\mathbf{k}_{3}\cdot\mathbf{r}}=\frac{H_{0}}{4}e^{i\left(k_{x}x-k_{z}z\right)}\mathbf{e}_{y},\nonumber\\\mathbf{H}_{4}&=&\frac{H_{0}}{4}e^{i\mathbf{k}_{4}\cdot\mathbf{r}}=\frac{H_{0}}{4}e^{-i\left(k_{x}x-k_{z}z\right)}\mathbf{e}_{y},
\end{eqnarray}
where $\mathbf{k}_{1}\cdot\mathbf{r}=-\mathbf{k}_{2}\cdot\mathbf{r}=k_{x}x+k_{z}z,$
$\mathbf{k}_{3}\cdot\mathbf{r}=-\mathbf{k}_{4}\cdot\mathbf{r}=k_{x}x-k_{z}z$,
and $k_{x}=k\mathrm{sin}\psi$, $k_{z}=k\mathrm{cos}\psi.$ Thus,
the total magnetic field at $\mathbf{r}=x\mathbf{e}_{x}+y\mathbf{e}_{y}+z\mathbf{e}_{z}$
can be written as
\begin{equation}
\mathbf{H}=  \frac{H_{0}}{2}\left[\mathrm{cos}\left(k_{x}x+k_{z}z+\phi\right)+\mathrm{cos}\left(k_{x}x-k_{z}z\right)\right]\mathbf{e}_{y},\nonumber
\end{equation}
and the corresponding electric field is given by
\begin{eqnarray}
\mathbf{E} & = & i\frac{E_{0}}{2}\mathrm{cos}\psi\left[\mathrm{sin}\left(k_{x}x+k_{z}z+\phi\right)-\mathrm{sin}\left(k_{x}x-k_{z}z\right)\right]\mathbf{e}_{x}\nonumber \\
 &  & -i\frac{E_{0}}{2}\mathrm{sin}\psi\left[\mathrm{sin}\left(k_{x}x+k_{z}z+\phi\right)+\mathrm{sin}\left(k_{x}x-k_{z}z\right)\right]\mathbf{e}_{z}.\nonumber
\end{eqnarray}
Using the above electric and magnetic fields and their derivatives, we can obtain the ensemble-averaged induced
multipole moments at the center of the cloud ($\mathbf{r}=0$)

\begin{eqnarray}
\left\langle \mathbf{d}^{\mathrm{E}}\right\rangle  & = & \varepsilon_{0}\left\langle \alpha_{_{\mathrm{ED}}}\right\rangle \mathbf{E}\left(\mathbf{r}=0\right)\nonumber\\
&=&\varepsilon_{0}\left\langle \alpha_{_{\mathrm{ED}}}\right\rangle iE_{0}\left(\mathrm{cos}\psi\mathbf{e}_{x}-\mathrm{sin}\psi\mathbf{e}_{z}\right)\mathrm{sin}\frac{\phi}{2}\mathrm{cos}\frac{\phi}{2},\nonumber \\
\left\langle \mathbf{d}^{\mathrm{M}}\right\rangle  & = & \left\langle \alpha_{_{\mathrm{MD}}}\right\rangle \mathbf{H}\left(\mathbf{r}=0\right)=\left\langle \alpha_{_{\mathrm{MD}}}\right\rangle H_{0}\mathrm{cos^{2}}\frac{\phi}{2},\nonumber \\
\left\langle \mathbf{Q}^{\mathrm{E}}\right\rangle  & = & \frac{1}{2}\varepsilon_{0}\left\langle \alpha_{_{\mathrm{EQ}}}\right\rangle \left.\left(\mathbf{\nabla\mathbf{E}}+\mathbf{\mathbf{E}\nabla}\right)\right|_{\mathbf{r}=0}\nonumber \\
 & = & \frac{1}{2}\varepsilon_{0}\left\langle \alpha_{_{\mathrm{EQ}}}\right\rangle iE_{0}k\mathrm{cos}2\psi\mathrm{cos^{2}}\frac{\phi}{2}\left(\mathbf{e}_{x}\mathbf{e}_{z}+\mathbf{e}_{z}\mathbf{e}_{x}\right)\nonumber \\
 &  & +\frac{1}{2}\varepsilon_{0}\left\langle \alpha_{_{\mathrm{EQ}}}\right\rangle iE_{0}k\mathrm{sin}2\psi\mathrm{sin^{2}}\frac{\phi}{2}\left(\mathbf{e}_{z}\mathbf{e}_{z}-\mathbf{e}_{x}\mathbf{e}_{x}\right),\nonumber \\
\left\langle \mathbf{Q}^{\mathrm{M}}\right\rangle  & = & \frac{1}{2}\left\langle \alpha_{_{\mathrm{MQ}}}\right\rangle \left.\left(\mathbf{\nabla H}+\mathbf{\mathbf{H}\nabla}\right)\right|_{\mathbf{r}=0}\label{eq:ME_ave_4PW_TM} \\
 & = &-\frac{H_{0}k}{2}\left\langle \alpha_{_{\mathrm{MQ}}}\right\rangle \mathrm{sin}\psi\mathrm{sin}\frac{\phi}{2}\mathrm{cos}\frac{\phi}{2}\left(\mathbf{e}_{y}\mathbf{e}_{x}+\mathbf{e}_{x}\mathbf{e}_{y}\right)\nonumber\\
 &&-\frac{H_{0}k}{2}\left\langle \alpha_{_{\mathrm{MQ}}}\right\rangle \mathrm{cos}\psi\left(\mathbf{e}_{y}\mathbf{e}_{z}+\mathbf{e}_{z}\mathbf{e}_{y}\right)\mathrm{sin}\frac{\phi}{2}\mathrm{cos}\frac{\phi}{2}.\nonumber
\end{eqnarray}
 Now by substituting Eq.~(\ref{eq:ME_ave_4PW_TM}) into Eqs.~(\ref{eq:C_sca_ME_coh}), we obtain
\begin{eqnarray}
C_{\mathrm{sca}}^{\mathrm{coh}} & = & \frac{3\lambda^{2}}{2\pi}\left|\left\langle \frac{\alpha_{_{\mathrm{ED}}}}{\alpha_{0}}\right\rangle \right|^{2}\mathrm{sin}^{2}\frac{\phi}{2}\mathrm{cos}^{2}\frac{\phi}{2}\nonumber\\
&&+\frac{3\lambda^{2}}{2\pi}\left|\left\langle \frac{\alpha_{_{\mathrm{MD}}}}{\alpha_{0}}\right\rangle \right|^{2}\mathrm{cos}^{4}\frac{\phi}{2}\nonumber \\
 &  & +\frac{5\lambda^{2}}{2\pi}\left|\left\langle \frac{\alpha_{_{\mathrm{EQ}}}}{\alpha_{0}}\right\rangle \right|^{2}\left[\mathrm{cos}^{2}2\psi\mathrm{cos}^{4}\frac{\phi}{2}+\mathrm{sin}^{2}2\psi\mathrm{sin}^{4}\frac{\phi}{2}\right]
 \nonumber \\
 &  & +\frac{5\lambda^{2}}{2\pi}\left|\left\langle \frac{\alpha_{_{\mathrm{MQ}}}}{\alpha_{0}}\right\rangle \right|^{2}\mathrm{sin}^{2}\frac{\phi}{2}\mathrm{cos}^{2}\frac{\phi}{2}.\label{eq:ME_4PW_TM}
\end{eqnarray}
And by substituting Eq.~(\ref{eq:ME_ave_4PW_TM}) into Eqs.~(\ref{eq:C_ext_ME_total}), the total (sum of incoherent and coherent) scattering (extinction)
cross section can be obtained
\begin{eqnarray}
C_{\mathrm{sca}}^{\mathrm{total}} & = & C_{\mathrm{ext}}=
 \frac{3\lambda^{2}}{2\pi}\mathrm{Im}\left[\left\langle \frac{\alpha_{_{\mathrm{ED}}}}{\alpha_{0}}\right\rangle \right]\mathrm{sin}^{2}\frac{\phi}{2}\mathrm{cos}^{2}\frac{\phi}{2}\nonumber\\
 &&+\frac{3\lambda^{2}}{2\pi}\mathrm{Im}\left[\left\langle \frac{\alpha_{_{\mathrm{MD}}}}{\alpha_{0}}\right\rangle \right]\mathrm{cos}^{4}\frac{\phi}{2}\nonumber \\
 &  & +\frac{5\lambda^{2}}{2\pi}\mathrm{Im}\left[\left\langle \frac{\alpha_{_{\mathrm{EQ}}}}{\alpha_{0}}\right\rangle \right]\left[\mathrm{cos}^{2}2\psi\mathrm{cos}^{4}\frac{\phi}{2}+\mathrm{sin}^{2}2\psi\mathrm{sin}^{4}\frac{\phi}{2}\right]\nonumber\\
 &&+\frac{5\lambda^{2}}{2\pi}\mathrm{Im}\left[\left\langle \frac{\alpha_{_{\mathrm{MQ}}}}{\alpha_{0}}\right\rangle \right]\mathrm{sin}^{2}\frac{\phi}{2}\mathrm{cos}^{2}\frac{\phi}{2}.\label{eq:C_ext_ME_4PW_TM}
\end{eqnarray}
Finally, in order to selectivity excite different multipole moments, we assume $\psi=\pi/4$ and consider two cases:

i) $\phi=2m\pi$: the induced moments read as
\begin{eqnarray}
\left\langle \mathbf{d}^{\mathrm{E}}\right\rangle  & = & 0,\,\,\,\,\left\langle \mathbf{Q}^{\mathrm{E}}\right\rangle =0,\,\,\,\,\,\left\langle \mathbf{Q}^{\mathrm{M}}\right\rangle =0,\nonumber \\
\left\langle \mathbf{d}^{\mathrm{M}}\right\rangle  & = & \left\langle \alpha_{_{\mathrm{MD}}}\right\rangle H_{0}\mathbf{e}_{y},
\end{eqnarray}
thus, \textit{only} the magnetic dipole moment is excited and the scattering
cross sections read as
\begin{eqnarray}
C_{\mathrm{sca}}^{\mathrm{coh}} & = & \frac{3\lambda^{2}}{2\pi}\left|\left\langle \frac{\alpha_{_{\mathrm{MD}}}}{\alpha_{0}}\right\rangle \right|^{2},\nonumber\\
C_{\mathrm{sca}}^{\mathrm{total}}& = &\frac{3\lambda^{2}}{2\pi}\mathrm{Im}\left[\left\langle \frac{\alpha_{_{\mathrm{MD}}}}{\alpha_{0}}\right\rangle \right].
\end{eqnarray}
ii) $\phi=(2m+1)\pi$: the induced moments read as
\begin{eqnarray}
\left\langle \mathbf{d}^{\mathrm{E}}\right\rangle  & = & 0,\,\,\,\,\,\left\langle \mathbf{d}^{\mathrm{M}}\right\rangle =0,\,\,\,\,\,\left\langle \mathbf{Q}^{\mathrm{M}}\right\rangle =0,\nonumber \\
\left\langle \mathbf{Q}^{\mathrm{E}}\right\rangle  & = & \frac{1}{2}\varepsilon_{0}\left\langle \alpha_{_{\mathrm{EQ}}}\right\rangle ikE_{0}\left(\mathbf{e}_{z}\mathbf{e}_{z}-\mathbf{e}_{x}\mathbf{e}_{x}\right),
\end{eqnarray}
thus, \textit{only} the electric quadrupole moment is excited and the scattering
cross sections read as
\begin{eqnarray}
C_{\mathrm{sca}}^{\mathrm{coh}} & = & \frac{5\lambda^{2}}{2\pi}\left|\left\langle \frac{\alpha_{_{\mathrm{EQ}}}}{\alpha_{0}^{\prime}}\right\rangle \right|^{2},\nonumber\\C_{\mathrm{sca}}^{\mathrm{total}}&=&\frac{5\lambda^{2}}{2\pi}\mathrm{Im}\left[\left\langle \frac{\alpha_{_{\mathrm{EQ}}}}{\alpha_{0}^{\prime}}\right\rangle \right].
\end{eqnarray}

\subsection{Four plane waves with TE polarization}
In this subsection, we consider an atomic cloud when illuminated by
four plane waves with TE polarization. The electric fields of the
plane waves are defined as
\begin{eqnarray}
\mathbf{E}_{1} & = & \frac{E_{0}}{4}e^{i\left(\mathbf{k}_{1}\cdot\mathbf{r}+\phi\right)}=\frac{E_{0}}{4}e^{i\left(k_{x}x+k_{z}z+\phi\right)}\mathbf{e}_{y},\nonumber\\
\mathbf{E}_{2}&=&\frac{E_{0}}{4}e^{i\left(\mathbf{k}_{2}\cdot\mathbf{r}-\phi\right)}=\frac{E_{0}}{4}e^{-i\left(k_{x}x+k_{z}z+\phi\right)}\mathbf{e}_{y},\nonumber \\
\mathbf{E}_{3} & = & \frac{E_{0}}{4}e^{i\mathbf{k}_{3}\cdot\mathbf{r}}=\frac{E_{0}}{4}e^{i\left(k_{x}x-k_{z}z\right)}\mathbf{e}_{y},\nonumber\\
\mathbf{E}_{4}&=&\frac{E_{0}}{4}e^{i\mathbf{k}_{4}\cdot\mathbf{r}}=\frac{E_{0}}{4}e^{-i\left(k_{x}x-k_{z}z\right)}\mathbf{e}_{y},
\end{eqnarray}
where $\mathbf{k}_{1}\cdot\mathbf{r}=-\mathbf{k}_{2}\cdot\mathbf{r}=k_{x}x+k_{z}z,$
$\mathbf{k}_{3}\cdot\mathbf{r}=-\mathbf{k}_{4}\cdot\mathbf{r}=k_{x}x-k_{z}z$,
and $k_{x}=k\mathrm{sin}\psi$, $k_{z}=k\mathrm{cos}\psi.$ Thus,
the total electric field at $\mathbf{r}=x\mathbf{e}_{x}+y\mathbf{e}_{y}+z\mathbf{e}_{z}$
can be written as
\begin{eqnarray}
\mathbf{E} & = & 
\frac{E_{0}}{2}\left[\mathrm{cos}\left(k_{x}x+k_{z}z+\phi\right)+\mathrm{cos}\left(k_{x}x-k_{z}z\right)\right]\mathbf{e}_{y},\nonumber
\end{eqnarray}
and the corresponding magnetic field is given by
\begin{eqnarray}
\mathbf{H} & = & -i\frac{H_{0}}{2}\mathrm{cos}\psi\left[\mathrm{sin}\left(k_{x}x+k_{z}z+\phi\right)-\mathrm{sin}\left(k_{x}x-k_{z}z\right)\right]\mathbf{e}_{x}\nonumber \\
 &  & +i\frac{H_{0}}{2}\mathrm{sin}\psi\left[\mathrm{sin}\left(k_{x}x+k_{z}z+\phi\right)+\mathrm{sin}\left(k_{x}x-k_{z}z\right)\right]\mathbf{e}_{z}.\nonumber
\end{eqnarray}

%%%%%%%%%%%%%%%%%%%%%% Table start
\begin{table*}
\caption{Selective excitation of subwavelength atomic clouds using superposition
of four plane waves. The first column shows the polarization and the relative phase of the waves. See Fig. 3 (a) of the main text for the geometry. The 2nd to 5th columns show the fields' amplitudes and their gradient at the center of the atomic cloud, based on which a particular multipole moment is excited as shown in the last column.
}

\resizebox{\textwidth}{!}{

\begin{tabular}{|c|c|c|c|c|cc|}
\hline 
 & \multicolumn{4}{c|}{%
\begin{tabular}{c}

\textbf{Fields and their gradients at the center of the atomic cloud}\tabularnewline
\tabularnewline
\end{tabular}} &  & \tabularnewline
\cline{2-5} \cline{3-5} \cline{4-5} \cline{5-5} 
\begin{tabular}{l}
\textbf{Plane waves' polarizations and phases}\tabularnewline
\tabularnewline
\tabularnewline
\end{tabular} & %
\begin{tabular}{l}
\tabularnewline
$\left.\mathbf{E}\right|_{\mathbf{r}=0}$\tabularnewline
\tabularnewline
\end{tabular} & %
\begin{tabular}{l}
\tabularnewline
$\left.\mathbf{H}\right|_{\mathbf{r}=0}$\tabularnewline
\tabularnewline
\end{tabular} & %
\begin{tabular}{l}
\tabularnewline
$\left.\left(\mathbf{\nabla\mathbf{E}}+\mathbf{\mathbf{E}\nabla}\right)\right|_{\mathbf{r}=0}$\tabularnewline
\tabularnewline
\end{tabular} & %
\begin{tabular}{l}
\tabularnewline
$\left.\left(\mathbf{\nabla H}+\mathbf{\mathbf{H}\nabla}\right)\right|_{\mathbf{r}=0}$\tabularnewline
\tabularnewline
\end{tabular} &  & %
\begin{tabular}{l}
\textbf{Pure excitation of induced multipoles}\tabularnewline
\tabularnewline
\tabularnewline
\end{tabular}\tabularnewline
\hline 
\begin{tabular}{l}
\tabularnewline
TE: ~$\phi=2m\pi$~~~~~~~~~~~\tabularnewline
\tabularnewline
\end{tabular} & %
\begin{tabular}{l}
\tabularnewline
$E_{0}\mathbf{e}_{y}$\tabularnewline
\tabularnewline
\end{tabular} & %
\begin{tabular}{l}
\tabularnewline
$0$\tabularnewline
\tabularnewline
\end{tabular} & %
\begin{tabular}{l}
\tabularnewline
$0$\tabularnewline
\tabularnewline
\end{tabular} & %
\begin{tabular}{l}
\tabularnewline
$0$\tabularnewline
\tabularnewline
\end{tabular} & %
\begin{tabular}{l}
\tabularnewline
ED:\tabularnewline
\tabularnewline
\end{tabular} & %
\begin{tabular}{l}
\tabularnewline
$\left\langle \mathbf{d}^{\mathrm{E}}\right\rangle =\varepsilon_{0}\left\langle \alpha_{_{\mathrm{ED}}}\right\rangle \left.\mathbf{E}\right|_{\mathbf{r}=0}$~~~~~~~~~~~~~~~~~~~\tabularnewline
\tabularnewline
\end{tabular}\tabularnewline
\hline 
\begin{tabular}{l}
\tabularnewline
TE: ~$\phi=\left(2m+1\right)\pi\,\,$\tabularnewline
\tabularnewline
\end{tabular} & %
\begin{tabular}{l}
\tabularnewline
$0$\tabularnewline
\tabularnewline
\end{tabular} & %
\begin{tabular}{l}
\tabularnewline
$0$\tabularnewline
\tabularnewline
\end{tabular} & %
\begin{tabular}{l}
\tabularnewline
$0$\tabularnewline
\tabularnewline
\end{tabular} & %
\begin{tabular}{l}
\tabularnewline
$ikH_{0}\left(\mathbf{e}_{x}\mathbf{e}_{x}-\mathbf{e}_{z}\mathbf{e}_{z}\right)$\tabularnewline
\tabularnewline
\end{tabular} & %
\begin{tabular}{l}
\tabularnewline
MQ:\tabularnewline
\tabularnewline
\end{tabular} & %
\begin{tabular}{l}
\tabularnewline
$\left\langle \mathbf{Q}^{\mathrm{M}}\right\rangle =\frac{1}{2}\left\langle \alpha_{_{\mathrm{MQ}}}\right\rangle \left.\left(\mathbf{\nabla H}+\mathbf{\mathbf{H}\nabla}\right)\right|_{\mathbf{r}=0}$~~~~~\tabularnewline
\tabularnewline
\end{tabular}\tabularnewline
\hline 
\begin{tabular}{l}
\tabularnewline
TM: ~$\phi=2m\pi$~~~~~~~~~~\tabularnewline
\tabularnewline
\end{tabular} & %
\begin{tabular}{l}
\tabularnewline
$0$\tabularnewline
\tabularnewline
\end{tabular} & %
\begin{tabular}{l}
\tabularnewline
$H_{0}\mathbf{e}_{y}$\tabularnewline
\tabularnewline
\end{tabular} & %
\begin{tabular}{l}
\tabularnewline
$0$\tabularnewline
\tabularnewline
\end{tabular} & %
\begin{tabular}{l}
\tabularnewline
$0$\tabularnewline
\tabularnewline
\end{tabular} & %
\begin{tabular}{l}
\tabularnewline
MD:\tabularnewline
\tabularnewline
\end{tabular} & %
\begin{tabular}{l}
\tabularnewline
$\left\langle \mathbf{d}^{\mathrm{M}}\right\rangle =\left\langle \alpha_{_{\mathrm{MD}}}\right\rangle \left.\mathbf{H}\right|_{\mathbf{r}=0}$~~~~~~~~~~~~~~~~~~~~~~\tabularnewline
\tabularnewline
\end{tabular}\tabularnewline
\hline 
\begin{tabular}{l}
\tabularnewline
TM: ~$\phi=\left(2m+1\right)\pi$\tabularnewline
\tabularnewline
\end{tabular} & %
\begin{tabular}{l}
\tabularnewline
$0$\tabularnewline
\tabularnewline
\end{tabular} & %
\begin{tabular}{l}
\tabularnewline
$0$\tabularnewline
\tabularnewline
\end{tabular} & %
\begin{tabular}{l}
\tabularnewline
$ikE_{0}\left(\mathbf{e}_{z}\mathbf{e}_{z}-\mathbf{e}_{x}\mathbf{e}_{x}\right)$\tabularnewline
\tabularnewline
\end{tabular} & %
\begin{tabular}{l}
\tabularnewline
$0$\tabularnewline
\tabularnewline
\end{tabular} & %
\begin{tabular}{l}
\tabularnewline
EQ:\tabularnewline
\tabularnewline
\end{tabular} & %
\begin{tabular}{l}
\tabularnewline
$\left\langle \mathbf{Q}^{\mathrm{E}}\right\rangle =\frac{1}{2}\varepsilon_{0}\left\langle \alpha_{_{\mathrm{EQ}}}\right\rangle \left.\left(\mathbf{\nabla\mathbf{E}}+\mathbf{\mathbf{E}\nabla}\right)\right|_{\mathbf{r}=0}$\tabularnewline
\tabularnewline
\end{tabular}\tabularnewline
\hline 
\end{tabular} \label{Table_Selective_Exciation}

}
\end{table*}

%%%%%%%%%%%%%%%%%%%%%%%%%%%%%%%%
Using the above electric and magnetic fields and their derivatives, we obtain the ensemble-averaged induced
multipole moments for four plane waves at the center of the cloud ($\mathbf{r}=0$)
\begin{eqnarray}
\left\langle \mathbf{d}^{\mathrm{E}}\right\rangle  & = & \varepsilon_{0}\left\langle \alpha_{_{\mathrm{ED}}}\right\rangle \mathbf{E}\left(\mathbf{r}=0\right)\nonumber\\
&=&\varepsilon_{0}\left\langle \alpha_{_{\mathrm{ED}}}\right\rangle E_{0}\mathrm{cos}^{2}\frac{\phi}{2}\mathbf{e}_{y},\nonumber \\
\left\langle \mathbf{d}^{\mathrm{M}}\right\rangle  & = & \left\langle \alpha_{_{\mathrm{MD}}}\right\rangle \mathbf{H}\left(\mathbf{r}=0\right)\nonumber\\
&=&i\left\langle \alpha_{_{\mathrm{MD}}}\right\rangle H_{0}\left(-\mathrm{cos}\psi\mathbf{e}_{x}+\mathrm{sin}\psi\mathbf{e}_{z}\right)\mathrm{sin}\frac{\phi}{2}\mathrm{cos}\frac{\phi}{2},\nonumber \\
\left\langle \mathbf{Q}^{\mathrm{E}}\right\rangle  & = & \frac{1}{2}\varepsilon_{0}\left\langle \alpha_{_{\mathrm{EQ}}}\right\rangle \left.\left(\mathbf{\nabla\mathbf{E}}+\mathbf{\mathbf{E}\nabla}\right)\right|_{\mathbf{r}=0}\nonumber \\
 &=&-\varepsilon_{0}\frac{E_{0}k}{2}\left\langle \alpha_{_{\mathrm{EQ}}}\right\rangle \mathrm{sin}\psi\mathrm{sin}\frac{\phi}{2}\mathrm{cos}\frac{\phi}{2}\left(\mathbf{e}_{y}\mathbf{e}_{x}+\mathbf{e}_{x}\mathbf{e}_{y}\right)\nonumber\\
 &&-\varepsilon_{0}\frac{E_{0}k}{2}\left\langle \alpha_{_{\mathrm{EQ}}}\right\rangle \mathrm{cos}\psi\mathrm{sin}\frac{\phi}{2}\mathrm{cos}\frac{\phi}{2}\left(\mathbf{e}_{y}\mathbf{e}_{z}+\mathbf{e}_{z}\mathbf{e}_{y}\right)
.\nonumber \\
\left\langle \mathbf{Q}^{\mathrm{M}}\right\rangle  & = & \frac{1}{2}\left\langle \alpha_{_{\mathrm{MQ}}}\right\rangle \left.\left(\mathbf{\nabla H}+\mathbf{\mathbf{H}\nabla}\right)\right|_{\mathbf{r}=0} \label{eq:ME_ave_4PW_TE}\\
 & = & -\frac{1}{2}\left\langle \alpha_{_{\mathrm{MQ}}}\right\rangle iH_{0}k\mathrm{cos}2\psi\mathrm{cos}^{2}\frac{\phi}{2}\left(\mathbf{e}_{x}\mathbf{e}_{z}+\mathbf{e}_{z}\mathbf{e}_{x}\right)\nonumber \\
 &  & +\frac{1}{2}\left\langle \alpha_{_{\mathrm{MQ}}}\right\rangle iH_{0}k\mathrm{sin}2\psi\mathrm{sin}^{2}\frac{\phi}{2}\left(\mathbf{e}_{x}\mathbf{e}_{x}-\mathbf{e}_{z}\mathbf{e}_{z}\right).\nonumber
\end{eqnarray}

Now by substituting Eq.~(\ref{eq:ME_ave_4PW_TE}) into Eq.~(\ref{eq:C_sca_ME_coh}), we obtain the coherent scattering cross section:
\begin{eqnarray}
C_{\mathrm{sca}}^{\mathrm{coh}} & = & \frac{3\lambda^{2}}{2\pi}\left|\left\langle \frac{\alpha_{_{\mathrm{ED}}}}{\alpha_{0}}\right\rangle \right|^{2}\mathrm{cos}^{4}\frac{\phi}{2}\label{eq:ME_4PW_TE}\\
&&+\frac{3\lambda^{2}}{2\pi}\left|\left\langle \frac{\alpha_{_{\mathrm{MD}}}}{\alpha_{0}}\right\rangle \right|^{2}\mathrm{sin}^{2}\frac{\phi}{2}\mathrm{cos}^{2}\frac{\phi}{2}\nonumber \\
 &  & +\frac{5\lambda^{2}}{2\pi}\left|\left\langle \frac{\alpha_{_{\mathrm{EQ}}}}{\alpha_{0}}\right\rangle \right|^{2}\mathrm{sin}^{2}\frac{\phi}{2}\mathrm{cos}^{2}\frac{\phi}{2}\nonumber\\
&&+\frac{5\lambda^{2}}{2\pi}\left|\left\langle \frac{\alpha_{_{\mathrm{MQ}}}}{\alpha_{0}}\right\rangle \right|^{2}\left[\mathrm{cos}^{2}2\psi\mathrm{cos}^{4}\frac{\phi}{2}+\mathrm{sin}^{2}2\psi\mathrm{sin}^{4}\frac{\phi}{2}\right].\nonumber
\end{eqnarray}

And by substituting Eq.~(\ref{eq:ME_ave_4PW_TE}) into Eq.~(\ref{eq:C_ext_ME_total}), the total scattering (or extinction) cross section can be obtained
\begin{eqnarray}
C_{\mathrm{sca}}^{\mathrm{total}} & = & C_{\mathrm{ext}}=
 \frac{3\lambda^{2}}{2\pi}\mathrm{Im}\left[\left\langle \frac{\alpha_{_{\mathrm{ED}}}}{\alpha_{0}}\right\rangle \right]\mathrm{cos}^{4}\frac{\phi}{2}\label{eq:C_ext_ME_4PW_TE}\\
&&+\frac{3\lambda^{2}}{2\pi}\mathrm{Im}\left[\left\langle \frac{\alpha_{_{\mathrm{MD}}}}{\alpha_{0}}\right\rangle \right]\mathrm{sin}^{2}\frac{\phi}{2}\mathrm{cos}^{2}\frac{\phi}{2}\nonumber \\
 &  & +\frac{5\lambda^{2}}{2\pi}\mathrm{Im}\left[\left\langle \frac{\alpha_{_{\mathrm{EQ}}}}{\alpha_{0}}\right\rangle \right]\mathrm{sin}^{2}\frac{\phi}{2}\mathrm{cos}^{2}\frac{\phi}{2}\nonumber\\
&&+\frac{5\lambda^{2}}{2\pi}\mathrm{Im}\left[\left\langle \frac{\alpha_{_{\mathrm{MQ}}}}{\alpha_{0}}\right\rangle \right]\left[\mathrm{cos}^{2}2\psi\mathrm{cos}^{4}\frac{\phi}{2}+\mathrm{sin}^{2}2\psi\mathrm{sin}^{4}\frac{\phi}{2}\right].\nonumber
\end{eqnarray}
Finally, in order to selectivity excite different multipole moments, we assume $\psi=\pi/4$ and consider two cases:

i) $\phi=2m\pi$: the induced moments read as
\begin{eqnarray}
\left\langle \mathbf{d}^{\mathrm{M}}\right\rangle  & = & 0,\,\,\,\,\,\left\langle \mathbf{Q}^{\mathrm{E}}\right\rangle =0,\,\,\,\,\,\left\langle \mathbf{Q}^{\mathrm{M}}\right\rangle =0,\nonumber\\
\left\langle \mathbf{d}^{\mathrm{E}}\right\rangle &=&\left\langle \alpha_{_{\mathrm{ED}}}\right\rangle E_{0}\mathbf{e}_{y},
\end{eqnarray}
thus \textit{only} the electric dipole moment is excited we excite and the scattering
cross sections read as
\begin{eqnarray}
C_{\mathrm{sca}}^{\mathrm{coh}} & = & \frac{3\lambda^{2}}{2\pi}\left|\left\langle \frac{\alpha_{_{\mathrm{ED}}}}{\alpha_{0}}\right\rangle \right|^{2},\nonumber\\
C_{\mathrm{sca}}^{\mathrm{total}}&=&\frac{3\lambda^{2}}{2\pi}\mathrm{Im}\left[\left\langle \frac{\alpha_{_{\mathrm{ED}}}}{\alpha_{0}}\right\rangle \right].
\end{eqnarray}

ii) $\phi=(2m+1)\pi$: the induced moments read as
\begin{eqnarray}
\left\langle \mathbf{d}^{\mathrm{E}}\right\rangle  & = & 0,\,\,\,\,\,\left\langle \mathbf{d}^{\mathrm{M}}\right\rangle =0,\,\,\,\,\,\left\langle \mathbf{Q}^{\mathrm{E}}\right\rangle =0,\nonumber \\
\left\langle \mathbf{Q}^{\mathrm{M}}\right\rangle  & = & \frac{1}{2}\left\langle \alpha_{_{\mathrm{MQ}}}\right\rangle ikH_{0}\left(\mathbf{e}_{x}\mathbf{e}_{x}-\mathbf{e}_{z}\mathbf{e}_{z}\right),
\end{eqnarray}
thus \textit{only} the magnetic quadrupole moment is excited and the scattering
cross sections read as
\begin{eqnarray}
C_{\mathrm{sca}}^{\mathrm{coh}} & = & \frac{5\lambda^{2}}{2\pi}\left|\left\langle \frac{\alpha_{_{\mathrm{MQ}}}}{\alpha_{0}^{\prime}}\right\rangle \right|^{2},\nonumber\\
C_{\mathrm{sca}}^{\mathrm{total}}&=&\frac{5\lambda^{2}}{2\pi}\mathrm{Im}\left[\left\langle \frac{\alpha_{_{\mathrm{MQ}}}}{\alpha_{0}^{\prime}}\right\rangle \right].
\end{eqnarray}

Table~\ref{Table_Selective_Exciation} presents a summary of selective excitation with four plane waves. It shows the fields amplitudes and their gradients at the center of the cloud for different polarizations and phases of four plane waves. The last column indicates which multipole moment is excited based on the field amplitudes and gradients at the center.

\end{document}